\documentclass[a4paper]{article}

\usepackage{jheppub}

\usepackage{amsmath}
\usepackage{amssymb}
\usepackage{bbm}
\usepackage{amsfonts}
\usepackage{tipa}
\usepackage{mathrsfs}

\usepackage{graphicx}

\title{Converting Classical Theories to Quantum Theories by Solutions of the Hamilton-Jacobi Equation}

\author{Zhi-Qiang Guo}
\author{and Iv\'{a}n Schmidt}
\emailAdd{zhiqiang.guo@usm.cl}
\emailAdd{ivan.schmidt@usm.cl}
\affiliation{Departamento de F\'{i}sica y Centro Cient\'{i}fico
Tecnol\'{o}gico de Valpara\'{i}so,\\ Universidad T\'{e}cnica Federico
Santa Mar\'{i}a,\\ Casilla 110-V, Valpara\'{i}so, Chile}

\abstract{By employing special solutions of the Hamilton-Jacobi equation and tools from lattice theories,
we suggest an approach to convert classical theories to quantum theories for mechanics and field theories. Some nontrivial results
are obtained for a gauge field and a fermion field. For a topologically massive gauge theory, we can obtain a first order Lagrangian with mass term.
For the fermion field, in order to make our approach feasible, we supplement the conventional Lagrangian with a surface term.
This surface term can also produce the massive term for the fermion.}


\begin{document}
\maketitle

\section{Introduction}\label{sec:1}

Quantum theories have achieved tremendous success in the passed century. There are two conventional approaches to convert classical theories to
quantum theories: canonical quantization and path integral quantization. These two approaches employ two basic objects in
classical theories: the Hamiltonian and the Lagrangian, which are two equivalent tools to formulate classical theories. However, besides these,
we know there is a third equivalent way to formulate classical theories: the Hamilton-Jacobi equation. In this paper, we show that
it is possible to derive quantum theories from classical theories by employing special solutions of the Hamilton-Jacobi equation,
together with some tools from lattice theories. This third approach turns out to be consistent with the path integral approach for most of the cases.
However, we also can obtain several new sectors for gauge fields and fermion fields. Specifically, we can obtain a massive Lagrangian of the first order for
the topological massive gauge theory introduced in~\cite{Deser:1982vy,Deser:1981wh}; While for a fermion field, we find that the mass of a fermion can be produced
by a surface term, which is a mass generating mechanism similar to the topological massive gauge theory.

This paper is organized as follows. In section \ref{sec:2}, starting with classical mechanics, we introduce the basic tools
and methodology used to derive quantum mechanics from special solutions of the Hamilton-Jacobi equation. We also introduce tools to find
solutions of the Hamilton-Jacobi equation for nonlinear theories in section \ref{sec:2}. We turn to scalar field theories in section \ref{sec:3}.
In this section, we introduce the covariant Hamilton-Jacobi equation for field theories, that is,
the De Donder-Weyl approach~\cite{DeDonder:1935,Weyl:1934xb,weyl1935geodesic} for field theories. Along with the discussions regarding mechanics,
we introduce tools for solving the De Donder-Weyl equation for nonlinear field theories, and stress the differences between mechanics and field theories.
In section \ref{sec:4} and section \ref{sec:5}, we discuss the topological massive gauge theories and fermion fields separately.
Section \ref{sec:6} is devoted to $\mathrm{SU}(2)$ Yang-Mills theories. We only obtain restricted solutions for Yang-Mills theories, and no firm
conclusions can be drawn from these restricted solutions. We provide further discussions and conclusions in sections \ref{sec:7} and \ref{sec:8}.

\section{Examples of Mechanics}\label{sec:2}

We discuss mechanics in this section. Take the harmonic oscillator, for example. The logic structure of these discussions is as follows:
First we present the path integral quantization for the harmonic oscillator and its lattice definition; Then we display its Hamilton-Jacobi equation and
find several solutions for the Hamilton-Jacobi equation. Based on the discussions above, we can find a close relation between the lattice definition
of path integral quantization and a special solution of the Hamilton-Jacobi equation. This relation will make it feasible to convert classical theories
to quantum theories by the special solution of the Hamilton-Jacobi equation.

\subsection{Linear Theories: Harmonic Oscillator}\label{sec:2.1}

\subsubsection{Path Integral Quantization for the Harmonic Oscillator}\label{sec:2.1.1}

For a massive harmonic oscillator in one dimension, its Lagrangian is given by
\begin{eqnarray}
\label{osi-lagrangian}
L(x,\dot{x})=\frac{1}{2}m{\dot{x}}^2-\frac{1}{2}m\omega^2 x^2,
\end{eqnarray}
where $\dot{x}=\frac{dx}{dt}$ is defined. For a time evolution from $t_a$ to $t_b$, the Green function or the Feynman kernel
is defined by the path integral in configuration space
\begin{eqnarray}
\label{osi-pathintegral}
K(x_b,t_b;x_a,t_a)=\int^{x(t_b)=x_b}_{x(t_a)=x_a}\mathcal{D}x(t)\mathcal{N}\mathrm{exp}^{~i S(t_b,t_a)},\hspace{4mm}
S(t_b,t_a)=\int^{t_b}_{t_a}L(x,\dot{x})dt,\nonumber
\end{eqnarray}
where $\mathcal{N}$ is a normalization factor. Here and hereafter we use natural units, so the Planck constant $\hbar=1$ is assumed.
Discretizing the interval $[t_a,~t_b]$ into $N$ equivalent smaller intervals with length $\epsilon=\frac{t_b-t_a}{N}$, that is,
let $t_a=t_0,~t_1=t_0+\epsilon,~\cdots,~t_k=t_0+k\epsilon,~\cdots,~t_b=t_N$, then the path integral (\ref{osi-pathintegral}) can be regarded as the limit
of multi-integrals and can be expressed as the following lattice version
\begin{eqnarray}
\label{osi-pathintegral-lattice}
K(x_b,t_b;x_a,t_a)=\lim_{\substack{N\rightarrow \infty \\{\epsilon}\rightarrow 0}}\prod^{N-1}_{j=1}\biggl[\int dx_j\Bigl(\frac{m}{2\pi i\epsilon }\Bigr)^{\frac{1}{2}}
\mathrm{exp}\biggl\{i\epsilon\sum^{N-1}_{j=1}\frac{m}{2}\Bigl[\Bigl(\frac{x_{j+1}-x_j}{\epsilon}\Bigr)^2-\omega^2\Bigl(\frac{x_{j+1}+x_j}{2}\Bigr)^2\Bigr]\biggl\}\biggr].
\end{eqnarray}

For infinitesimal time evolution $\epsilon$, we have the short-time Feynman kernel
\begin{eqnarray}
\label{short-time-feyn-kernel}
K(x_{j+1},t_j+\epsilon;x_j,t_j)
=\Bigl(\frac{m}{2\pi i\epsilon }\Bigr)^{\frac{1}{2}}\mathrm{exp}\biggl\{i\epsilon\frac{m}{2}\Bigl[\Bigl(\frac{x_{j+1}-x_j}
{\epsilon}\Bigr)^2-\omega^2\Bigl(\frac{x_{j+1}+x_j}{2}\Bigr)^2\Bigr]\biggl\}.
\end{eqnarray}

So the Feynman kernel (\ref{osi-pathintegral-lattice}) of finite time evolution also can be regarded as multi-convolutions of short-time
Feynman kernels (\ref{short-time-feyn-kernel})
\begin{eqnarray}
\label{osi-pathintegral-lattice-feynman-kernel}
K(x_b,t_b;x_a,t_a)=\lim_{\substack{N\rightarrow \infty \\ \epsilon\rightarrow 0}} \int\int\cdots\int dx_{N-1}
\cdots dx_2 dx_1\hspace{52mm} \nonumber\\
K(x_b,t_{N-1}+\epsilon;x_{N-1},t_{N-1})\cdots K(x_2,t_1+\epsilon;x_1,t_1)K(x_1,t_a+\epsilon;x_a,t_a).
\end{eqnarray}
The Gaussian integrals in (\ref{osi-pathintegral-lattice}) can be performed in sequence and we can get a closed formulation for the Feynman kernel
\begin{eqnarray}
\label{osi-feyn-kernel-result}
K(x_b,t_b;x_a,t_a)\hspace{0mm}
=\Bigl(\frac{m\omega}{2\pi i\sin{\omega(t_b-t_a)}}\Bigr)^{\frac{1}{2}}
\mathrm{exp}\biggl\{\frac{im\omega}{2\sin{\omega(t_b-t_a)}}
[\cos{\omega(t_b-t_a)}(x^2_b+x^2_a)-2 x_b x_a]\biggr\}.
\end{eqnarray}

\subsubsection{Hamilton-Jacobi Equation for the Harmonic Oscillator}\label{sec:2.1.2}

The Hamilton-Jacobi equation was independently introduced by Hamilton and Jacobi from different approaches. In this section, we only give a
pedagogical introduction based on the independent integral of Hilbert \cite{Courant:1962methods},
which has been applied to field theories by De Donder \cite{DeDonder:1935} and Weyl \cite{Weyl:1934xb,weyl1935geodesic}. We caution that our introductions only
work well for regular Lagrangians, which we always work with in this paper. For non-regular Lagrangians, we refer the reader to the more rigorous discussions
in \cite{Weyl:1934xb,weyl1935geodesic,Rund:1935,Kastrup:1982qq,Gotay:1997eg}.

For a system of mechanics in one dimension, its classical aspects can be formulated by a Lagrangian $L\Bigl(q(t),\dot{q}(t),t\Bigr)$, where $q(t)$ is the
coordinate and $\dot{q}=\frac{dq}{dt}$ is defined. After performing the Legendre transformation $p=\frac{\partial  L}{\partial q}$, we can get the
corresponding Hamiltonian $H(p,q,t)=\dot{q}p-L$. The Hamilton-Jacobi equation can be derived as follows. Suppose there is a function $S(q,t)$ that depends only the coordinate but not its time derivative $\dot{q}(t)$, then the Lagrangian can be regarded as an independent integral of Hilbert with the following meaning
\begin{eqnarray}
\label{Hilbert-indep}
L=\frac{d S}{d t}=\frac{\partial  S}{\partial{t}}+\frac{dq}{dt}\frac{\partial{S}}{\partial{q}},
\end{eqnarray}
which also can be expressed as
\begin{eqnarray}
\label{Hilbert-indep-mod}
\frac{\partial  S}{\partial t}+\frac{dq}{dt}\frac{\partial{S}}{\partial{q}}-L=0.
\end{eqnarray}
Designating $p=\frac{\partial{S}}{\partial{q}}$, we can get the Hamilton-Jacobi equation straightforwardly
\begin{eqnarray}
\label{mechanics-hj}
\frac{\partial{S}}{\partial{t}}+\frac{dq}{dt}\frac{\partial{S}}{\partial{q}}-L=\frac{\partial{S}}{\partial{t}}+\frac{dq}{dt}p-L
=\frac{\partial{S}}{\partial{t}}+H\left(q,\frac{\partial{S}}{\partial{q}},t\right)=0.
\end{eqnarray}

For the harmonic oscillator (\ref{osi-lagrangian}), its Hamiltonian is given by
\begin{eqnarray}
\label{Hilbert-indep}
H=\frac{1}{2 m}p^2+\frac{1}{2}m^2 \omega^2 x^2.
\end{eqnarray}
Designating $p=\frac{\partial  S}{\partial x}$, we derive its Hamilton-Jacobi from eq.~(\ref{mechanics-hj}) as
\begin{eqnarray}
\label{osci-hj}
\frac{\partial S}{\partial t}+\frac{1}{2 m}\Bigl(\frac{\partial  S}{\partial x}\Bigr)^2+\frac{1}{2}m \omega^2 x^2=0.
\end{eqnarray}
We can find solutions of eq.~(\ref{osci-hj}) following two different approaches as follows:

Type-(I): Assuming the solution is a polynomial of $x$, we have
\begin{eqnarray}
\label{osci-hj-solution-assumption-poly}
S=\frac{1}{2}f(t)x^2+h(t)x+g(t).
\end{eqnarray}
Substituting this assumption into eq.~(\ref{osci-hj}), and letting the coefficients of $x$ to be zeros, we can get equations
\begin{eqnarray}
\label{osci-hj-solution-assumption-poly-co-2}
\frac{1}{2}\frac{d f(t)}{dt}+\frac{1}{2m}f(t)^2+\frac{1}{2}m\omega^2&=&0,\\
\label{osci-hj-solution-assumption-poly-co-1}
\frac{d h(t)}{dt}+\frac{1}{m}f(t)h(t)&=&0,\\
\label{osci-hj-solution-assumption-poly-co-0}
\frac{d g(t)}{dt}+\frac{1}{2m}h(t)h(t)&=&0.
\end{eqnarray}
Solving these ordinary differential equations~(ODEs), we can get the solution for $S$ to be
\begin{eqnarray}
\label{osci-hj-solution-poly}
(\mathrm{Ia}):\hspace{4mm}S=\frac{m}{2}\frac{\omega}{\sin{\omega(t-t_0)}}[\cos{\omega(t-t_0)}(x^2+x^2_0)-2 x x_0],\\
\label{osci-hj-solution-poly-1}
(\mathrm{Ib}):\hspace{2mm}S_1=\frac{m}{2}\frac{\omega}{\cos{\omega(t-t_0)}}[-\sin{\omega(t-t_0)}(x^2+x^2_0)-2 x x_0],
\end{eqnarray}
where $x_0,~t_0$ are integration constants.

Type-(II): Assuming that
\begin{eqnarray}
\label{osci-hj-solution-assumption-poly-non}
\mathcal{S}=-E(t-t_0)+W(x),
\end{eqnarray}
we can get a solution which we are familiar with in classical mechanics
\begin{eqnarray}
\label{osci-hj-solution-poly-non}
\mathcal{S}=-E(t-t_0)+\frac{E}{m}\arctan{\Bigl(\frac{m\omega x}{\sqrt{2mE-m^2\omega^2x^2}}\Bigr)}
+\frac{1}{2}x\sqrt{2mE-m^2\omega^2x^2},
\end{eqnarray}
where $E,~t_0$ are constants of integral.

\subsubsection{The Pauli Short-Time Kernel}\label{sec:2.1.3}

Based on the discussions above, we should have noticed the resemblance between the Feynman kernel (\ref{osi-feyn-kernel-result}) and the
solution (\ref{osci-hj-solution-poly}) of the Hamilton-Jacobi equation. Of course, this resemblance is not so surprising. The reasons are as follows:
For a Lagrangian of quadratic interaction, its Feynman kernel can be given exactly by the Wentzel-Kramers-Brillouin semiclassical approximation;
While the Wentzel-Kramers-Brillouin approximation involves the classical action of the system. Therefore, the resemblance between (\ref{osi-feyn-kernel-result}) and
(\ref{osci-hj-solution-poly}) only holds for quadratic interaction or for linear theories. For nonlinear theories, there would be no such resemblance.
However, we will show that a connection between the discretized Lagrangian and special solutions of the Hamilton-Jacobi equation can still be constructed,
not only for linear theories but also for nonlinear theories. In quantum mechanics, the link connecting them is Pauli's formula or
Pauli short-time kernel \cite{Pauli:1973,Morette:1951zz}.

For infinitesimal time evolution from $t$ to $t+\epsilon$, the Pauli short-time kernel is defined by
\begin{eqnarray}
\label{pauli-kernel}
K_P(q,t+\epsilon;q^\prime,t)=\Bigl(\frac{1}{2\pi i}\Bigr)^{\frac{1}{2}}\biggl(-\frac{\partial^2 S(q,q^\prime;\epsilon)}
{\partial q\partial q^\prime }\biggr)^{\frac{1}{2}}\mathrm{exp}^{i S(q,q^\prime;\epsilon)}.
\end{eqnarray}
Pauli proved that this kernel will satisfy the same equation as the Feynman short-time kernel (\ref{short-time-feyn-kernel}) if the
function $S(q,q^\prime;\epsilon)$ satisfies the Hamilton-Jacobi equation
\begin{eqnarray}
\label{mechanics-hj-pauli}
\frac{\partial S(q,q^\prime;\epsilon)}{\partial \epsilon}+H(q,\frac{\partial S}{\partial q})=0.
\end{eqnarray}
For the harmonic oscillator, the solutions of its Hamilton-Jacobi equation are given by eqs.~(\ref{osci-hj-solution-poly}), (\ref{osci-hj-solution-poly-1}) and (\ref{osci-hj-solution-poly-non}). Obviously the Type-(Ia) solution eq.~(\ref{osci-hj-solution-poly}) is suitable to formulate the Pauli short-time kernel while the solutions
Type-(Ib) (\ref{osci-hj-solution-poly-1}) and Type-(II) (\ref{osci-hj-solution-poly-non}) are not.
The Pauli short-time kernel for the harmonic oscillator is therefore given by
\begin{eqnarray}
\label{osi-pauli-kernel-short}
K_P(x_{j+1},t_{j+1};x_j,t_j)
=\Bigl(\frac{m\omega}{2\pi i\sin{\omega\epsilon}}\Bigr)^{\frac{1}{2}}\mathrm{exp}\biggl\{i\frac{m}{2}
\frac{\omega}{\sin{\omega\epsilon}}\Bigl[\cos{\omega\epsilon}(x^2_{j+1}+x^2_j)-2 x_{j+1} x_j\Bigr]\biggr\}.
\end{eqnarray}
Here we use the subscript $P$ to differentiate the Pauli short-time kernel from the Feynman short-time kernel.
The convolution of the Pauli short-time kernel (\ref{osi-pauli-kernel-short}) has the semi-group property, that is,
\begin{eqnarray}
\label{osi-pauli-kernel-short-semigroup}
K_P(x_{j+2},t_j+2\epsilon;x_j,t_j)
=\int dx_{j+1}K_P(x_{j+2},t_{j+1}+\epsilon;x_{j+1},t_{j+1})K_P(x_{j+1},t_j+\epsilon;x_j,t_j).
\end{eqnarray}
So its multi-convolutions will give the same results as the Feynman kernel (\ref{osi-feyn-kernel-result}).
The calculations of the multi-convolutions of the Pauli short-time kernel (\ref{osi-pauli-kernel-short}) are simple and straightforward;
While the multi-convolutions of the Feynman short-time kernel (\ref{short-time-feyn-kernel}) are very complicated.

\subsubsection{Deriving Discretized Lagrangians from Special Solutions of the Hamilton-Jacobi Equation}\label{sec:2.1.4}

The discussions in section~\ref{sec:2.1.3} suggest us an approach to convert classical theories to quantum theories.
The logic is as follows: Employing the multi-convolutions of the Pauli short-time kernel (\ref{osi-pauli-kernel-short}),
we can get the same results as that we get with the Feynman short-time kernel (\ref{short-time-feyn-kernel});
While the Pauli short-time kernel (\ref{osi-pauli-kernel-short}) can be determined by the special Type-(Ia) solution eq.~(\ref{osci-hj-solution-poly})
of the Hamilton-Jacobi equation. The logical deductions above can be reversed: we can begin with the Type-(Ia) solution eq.~(\ref{osci-hj-solution-poly}),
then we can get a discretized version of the path integral via the multi-convolutions of Pauli's formula (\ref{pauli-kernel}).

The equivalence of the results of these two approaches can be understood in another way. Expanding the Pauli short-time
kernel (\ref{osi-pauli-kernel-short}) with the small parameters $\epsilon$ and keeping the terms of leading order, we get
\begin{eqnarray}
\label{osi-pauli-kernel-short-expand-equ}
2\left[\cos{\omega\epsilon}(x^2_{j+1}+x^2_j)-2 x_{j+1} x_j\right]
=(1+\cos{\omega\epsilon})(x_{j+1}-x_j)^2-(1-\cos{\omega\epsilon})(x_{j+1}+x_j)^2\hspace{10mm}\\
\label{osi-pauli-kernel-short-expand-factor}
\Bigl(\frac{m\omega}{2\pi i\sin{\omega\epsilon}}\Bigr)^{\frac{1}{2}}\xrightarrow{\epsilon\rightarrow 0}
\Bigl(\frac{m}{2\pi i\epsilon }\Bigr)^{\frac{1}{2}}\hspace{60mm}\\
\label{osi-pauli-kernel-short-expand}
\mathrm{exp}\biggl\{i\frac{m}{2}
\frac{\omega}{\sin{\omega\epsilon}}\Bigl[\cos{\omega\epsilon}(x^2_{j+1}+x^2_j)-2 x_{j+1} x_j\Bigr]\biggr\}
\xrightarrow{\epsilon\rightarrow 0}\mathrm{exp}\biggl\{i\epsilon\Bigl[\frac{m}{2}\Bigl(\frac{x_{j+1}-x_j}{\epsilon}\Bigr)^2
-\frac{1}{2}m\omega^2\Bigl(\frac{x_{j+1}+x_j}{2}\Bigr)^2\Bigr]\biggl\}.\nonumber\\
\end{eqnarray}
What we should realize from this approximation is the discretized Lagrangian in the lattice definition (\ref{osi-pathintegral-lattice}) of the path integral is recovered from the Type-(Ia) solution eq.~(\ref{osci-hj-solution-poly}) of the Hamilton-Jacobi equation in the approximation of small $\epsilon$.

The derivation of the discretized Lagrangian can be generally summarized in the following procedure.
For a solution of the Hamilton-Jacobi equation $S(x,t;x_0,t_0)$ with some constants $(x_0,t_0)$ which need to be designated
by the initial conditions and boundary conditions, its corresponding discretized Lagrangian can be defined by
\begin{eqnarray}
\label{dis-lag-sol-hj-def}
\tilde{L}_{\mathrm{lattice}}=\frac{S(x_b,t_b;x_0,t_0)-S(x_a,t_a;x_0,t_0)}{t_b-t_a}.
\end{eqnarray}
Here $[t_a,t_b]$ is the lattice interval. The constants $(x_0,t_0)$ need to be designated appropriately.
Substituting the Type-(Ia) solution eq.~(\ref{osci-hj-solution-poly}) into eq.~(\ref{dis-lag-sol-hj-def}), we get
\begin{eqnarray}
\label{dis-lag-sol-hj-osci-def}
\tilde{L}_{\mathrm{lattice}}&=&\frac{1}{t_b-t_a}
\biggl\{\frac{m}{2}\frac{\omega}{\sin{\omega(t_b-t_0)}}[\cos{\omega(t_b-t_0)}(x_b^2+x^2_0)-2 x_b x_0]\hspace{0mm}\nonumber\\
&-&\frac{m}{2}\frac{\omega}{\sin{\omega(t_a-t_0)}}[\cos{\omega(t_a-t_0)}(x_a^2+x^2_0)-2 x_a x_0]\biggl\}.\hspace{0mm}
\end{eqnarray}
Taking the limit $x_0\rightarrow x_a$, then the limit $t_0\rightarrow t_a$, we get
\begin{eqnarray}
\label{dis-lag-sol-hj-osci}
L_{\mathrm{lattice}}=\lim_{\substack{t_0\rightarrow t_a}}\lim_{\substack{x_0\rightarrow x_a}}\tilde{L}_{\mathrm{lattice}}=\frac{1}{t_b-t_a}
\biggl\{\frac{m}{2}\frac{\omega}{\sin{\omega(t_b-t_a)}}[\cos{\omega(t_b-t_a)}(x_b^2+x^2_a)-2 x_b x_a]\biggl\}.\nonumber
\end{eqnarray}
For infinitesimal lattice spacing $t_b-t_a=\epsilon$, following the procedure in eq.~(\ref{osi-pauli-kernel-short-expand}),
eq.~(\ref{dis-lag-sol-hj-osci}) can be approximated as
\begin{eqnarray}
\label{dis-lag-sol-hj-osci-lattice}
L_{\mathrm{lattice}}=\frac{1}{t_b-t_a}\biggl\{\frac{m}{2}\frac{\omega}{\sin{\omega(t_b-t_a)}}
[\cos{\omega(t_b-t_a)}(x_b^2+x^2_a)-2 x_b x_a]\biggl\}\hspace{0mm}\nonumber\\
\xrightarrow{t_b-t_a\rightarrow 0}\frac{m}{2}\Bigl(\frac{x_{b}-x_a}{t_b-t_a}\Bigr)^2-\frac{1}{2}m\omega^2\Bigl(\frac{x_b+x_a}{2}\Bigr)^2.\hspace{30mm}
\end{eqnarray}
The lattice Lagrangian is recovered again.

\subsection{Nonlinear Theories: Double-Well Potential}\label{sec:2.2}
In this section, taking the double-well potential for example, we deal with nonlinear theories. For nonlinear theories, the Hamilton-Jacobi equation
is generally difficult to solve. We introduce the tools to handle the nonlinear theories and show that the discussions in section~(\ref{sec:2.1}) can
also apply to nonlinear theories.

\subsubsection{Tools for Solving the Nonlinear Hamilton-Jacobi Equation}\label{sec:2.2.1}

For a particle in the double-well potential, its Lagrangian is given by
\begin{eqnarray}
\label{well-lag}
L(x,\dot{x},t)=\frac{1}{2}m\dot{x}^2-\frac{1}{8}g^2\left(x^2-v^2\right)^2,
\end{eqnarray}
where $v$ is constant. After Legendre transformation $p=\frac{\partial L}{\partial \dot{x}}$, we get its Hamiltonian
\begin{eqnarray}
\label{well-ham}
H(x,p,t)=\frac{1}{2m}p^2+\frac{1}{8}g^2\left(x^2-v^2\right)^2.
\end{eqnarray}
The canonical Hamiltonian equations of motion are
\begin{eqnarray}
\label{well-ham-can-x}
\dot{x}&=&\frac{\partial H}{\partial p}=\frac{p}{m},\\
\label{well-ham-can-p}
\dot{p}&=&-\frac{\partial H}{\partial x}=-\frac{1}{2}g^2 x\left(x^2-v^2\right).
\end{eqnarray}
While the corresponding Euler-Lagrange equation of motion is
\begin{eqnarray}
\label{well-euler-lag}
m \ddot{x}+\frac{1}{2}g^2x\left(x^2-v^2\right)=0.
\end{eqnarray}
Designating $p=\frac{\partial S}{\partial x}$, its Hamilton-Jacobi equation is
\begin{eqnarray}
\label{well-ham-jacobi}
\frac{\partial S}{\partial t}+\frac{1}{2m}\left(\frac{\partial S}{\partial x}\right)^2+\frac{1}{8}g^2\left(x^2-v^2\right)^2=0.
\end{eqnarray}
This nonlinear equation is difficult to solve. However, employing the \lq\lq{embedding} method\rq\rq~introduced in \cite{vonRieth:1982ac}, we can
find a series solution for eq.~(\ref{well-ham-jacobi}); While this series solution is enough to satisfy our purpose in this paper.

The \lq\lq{embedding} method\rq\rq~is as follows. Suppose that we seek a series solution of the type
\begin{eqnarray}
\label{well-ham-jacobi-sol-series}
S(x,t)&=&S^{\ast}(t)+A(t)\left[x-f(t)\right]+R(t)\left[x-f(t)\right]^2\nonumber\\
&+&K(t)\left[x-f(t)\right]^3+M(t)\left[x-f(t)\right]^4+N(t)\left[x-f(t)\right]^5+\cdots,
\end{eqnarray}
where $f(t)$ is a function which will be given later. So it seems like we expand $S(x,t)$ in a series
around a function $f(t)$ in eq.~(\ref{well-ham-jacobi-sol-series}).
The potential function $V(x)=\frac{1}{8}g^2\left(x^2-v^2\right)^2$ is a polynomial, which can be expanded by the following identities
\begin{eqnarray}
\label{well-poten-expan}
V(x)=\frac{1}{8}g^2(x^2-v^2)^2
&=&\frac{1}{8}g^2\left[\left(x-f(t)+f(t)\right)^2-v^2\right]^2,\\
\left(x-f(t)+f(t)\right)^2&=&\left[x-f(t)\right]^2+2f(t)\left[x-f(t)\right]+f(t)^2.\nonumber
\end{eqnarray}
Notice that we expand these functionals around a function $f(t)$ but not $0$, so the combination $[x-f(t)]$ always remains.
Substituting the expansions (\ref{well-ham-jacobi-sol-series}) and (\ref{well-poten-expan}) into the Hamilton-Jacobi equation (\ref{well-ham-jacobi}),
collecting the terms of $[x-f(t)]$ of the same power, and letting the coefficients of this series to be zeros, we get
%\begin{widetext}
\begin{eqnarray}
\label{well-hj-sol-co-0}
\Bigl[x-f(t)\Bigr]^0:\hspace{19mm}\frac{d S^{\ast}(t)}{dt}+\frac{1}{2m}A^2+\frac{1}{8}g^2\left[f(t)^2-v^2\right]^2-A\frac{d f(t)}{dt}&=&0,\\
\label{well-hj-sol-co-1}
\Bigl[x-f(t)\Bigr]:\hspace{17mm}\frac{d A(t)}{dt}-2R\Bigl(\frac{d f(t)}{dt}-\frac{A}{m}\Bigr)+\frac{1}{2}g^2f(t)\left[f(t)^2-v^2\right]&=&0,\\
\label{well-hj-sol-co-2}
\Bigl[x-f(t)\Bigr]^2:\hspace{7mm}\frac{d R(t)}{dt}-3H\Bigl(\frac{d f(t)}{dt}-\frac{A}{m}\Bigr)+\frac{2}{m}R^2+\frac{1}{4}g^2\left[3f(t)^2-v^2\right]&=&0,\\
\label{well-hj-sol-co-3}
\Bigl[x-f(t)\Bigr]^3:\hspace{18.5mm}\frac{dK(t)}{dt}-4M\Bigl(\frac{d f(t)}{dt}-\frac{A}{m}\Bigr)+\frac{6}{m}KR+\frac{1}{2}g^2f(t)&=&0,\\
\label{well-hj-sol-co-4}
\Bigl[x-f(t)\Bigr]^4:\hspace{4mm}\frac{d M(t)}{dt}-5N\Bigl(\frac{d f(t)}{dt}-\frac{A}{m}\Bigr)+\frac{1}{2m}\left(9K^2+16MR\right)+\frac{1}{8}g^2&=&0,
\end{eqnarray}
%\end{widetext}
where we only display terms of power not higher than 4 while those of higher power are omitted for convenience. These ODEs are generally difficult to solve. The key ingredient of the \lq\lq{embedding} method\rq\rq~ is we can use some assumptions to simplify these ODEs. We suppose
\begin{eqnarray}
\label{well-hj-sol-co-1-sup-1}
\frac{d f(t)}{dt}-\frac{A}{m}&=&0,\\
\label{well-hj-sol-co-1-sup-2}
\frac{d A(t)}{dt}+\frac{1}{2}g^2f(t)\left[f(t)^2-v^2\right]&=&0,
\end{eqnarray}
then eq.~(\ref{well-hj-sol-co-1}) is satisfied. We noticed that equations (\ref{well-hj-sol-co-1-sup-1}) and (\ref{well-hj-sol-co-1-sup-2}) coincide
with the canonical Hamiltonian equations (\ref{well-ham-can-x}) and (\ref{well-ham-can-p}) if we make the replacements $x\rightarrow f(t)$ and $p\rightarrow A$.
So $f(t)$ satisfies the Euler-Lagrange equation automatically
\begin{eqnarray}
\label{well-euler-lag-f}
m \ddot{f}+\frac{1}{2}g^2f\left(f^2-v^2\right)=0.
\end{eqnarray}
Eq.~(\ref{well-hj-sol-co-0}) can be transformed as
\begin{eqnarray}
\label{well-hj-sol-co-0-sim}
\frac{d S^{\ast}(t)}{dt}&=&A\frac{d f(t)}{dt}-\frac{1}{2m}A^2-\frac{1}{8}g^2\left[f(t)^2-v^2\right]^2=L^{\ast}(f,\dot{f},t).
\end{eqnarray}
We see under the assumptions (\ref{well-hj-sol-co-1-sup-1}) and (\ref{well-hj-sol-co-1-sup-2}), the righthand side of (\ref{well-hj-sol-co-0-sim}) coincides
with the Lagrangian evaluated on the configuration $f(t)$. While eqs.~(\ref{well-hj-sol-co-2}), (\ref{well-hj-sol-co-3}) and (\ref{well-hj-sol-co-4}) are
simplified to be
\begin{eqnarray}
\label{well-hj-sol-co-2-sim}
\frac{d R(t)}{dt}+\frac{2}{m}R^2+\frac{1}{4}g^2\left[3f(t)^2-v^2\right]&=&0,\\
\label{well-hj-sol-co-3-sim}
\frac{dK(t)}{dt}+\frac{6}{m}KR+\frac{1}{2}g^2f(t)&=&0,\\
\label{well-hj-sol-co-4-sim}
\frac{d M(t)}{dt}+\frac{1}{2m}\left(9K^2+16MR\right)+\frac{1}{8}g^2&=&0.
\end{eqnarray}
We see that eq.~(\ref{well-hj-sol-co-2-sim}) only includes coefficients of power less than 3, eq.~(\ref{well-hj-sol-co-3-sim}) only includes coefficients of
power less than 4, and eq.~(\ref{well-hj-sol-co-4-sim}) only includes coefficients of power less than 5. So coefficients of higher power terms decouple
from that of lower power terms. This is a great advantage of the \lq\lq{embedding} method\rq\rq. From the discussions above,
we learned an important point in the \lq\lq{embedding} method\rq\rq~, which is we need to know a solution for the original Hamiltonian canonical equations
or the Euler-Lagrange equation. This solution of the equation of motion will help us find a series solution for the Hamilton-Jacobi equation. Obviously, this \lq\lq{embedding} method\rq\rq~is useless if our purpose is to find the solution of Hamiltonian canonical equations. However, here the \lq\lq{embedding} method\rq\rq~just satisfies our purpose because we try to find a solution of the Hamilton-Jacobi equation.

A solution of eq.~(\ref{well-euler-lag-f}) can be expressed by the Jacobi's function
\begin{eqnarray}
\label{well-euler-lag-f-sol}
f(t)=x_0\mathrm{JacobiDN}\Bigl(\lambda(t-t_0),k\Bigr),\hspace{2mm}
\lambda=\frac{gx_0}{2\sqrt{m}},\hspace{2mm}k^2=2\biggl(1-\Bigl(\frac{v}{x_0}\Bigr)^2\biggr),
\end{eqnarray}
where we have used the initial condition $f(t_0)=x_0$. Substituting the expression of $f(t)$
into eqs.~(\ref{well-hj-sol-co-2-sim}), (\ref{well-hj-sol-co-3-sim}) and (\ref{well-hj-sol-co-4-sim}), we can obtain solutions for $R(t)$, $K(t)$ and $M(t)$.
However, because of the limiting procedures in eqs.~(\ref{dis-lag-sol-hj-osci}) and (\ref{dis-lag-sol-hj-osci-lattice}), the solutions
for small $(t-t_0)$ are enough for our purpose. For small $(t-t_0)$, $f(t)$ can be replaced by $x_0$ in
eqs.~(\ref{well-hj-sol-co-0-sim}), (\ref{well-hj-sol-co-2-sim}), (\ref{well-hj-sol-co-3-sim}) and (\ref{well-hj-sol-co-4-sim}).
The solutions for $S^{\ast}(t)$, $R(t)$, $K(t)$ and $M(t)$ can be given by
\begin{eqnarray}
\label{well-hj-sol-co-0-sim-sol}
S^{\ast}(t)&=&-\frac{1}{8}g^2\Bigl(x_0^2-v^2\Bigr)^2r+O(r^2),\\
\label{well-hj-sol-co-2-sim-sol}
R(t)&=&\frac{m}{2}\frac{1}r-\frac{m}{6}\omega r+O(r^3),\\
\label{well-hj-sol-co-3-sim-sol}
K(t)&=&\frac{C_2}{r^3}+\frac{C_2}{2}\frac{\omega}{r}+\frac{C_1}{8}\omega^2 r+O(r^3),\\
\label{well-hj-sol-co-4-sim-sol}
M(t)&=&\frac{9}{2m}\frac{C^2_2}{r^5}+\frac{C_3}{r^4}+\frac{3\omega}{2m}\frac{C^2_2}{r^3}+\frac{2\omega}{3}\frac{C_3}{r^2}
+\frac{\omega^2}{24m}\Bigl(-9C_1+7C_2\Bigr)C_2\frac{1}{r}+\frac{2\omega^2}{9}C_3\\
&-&\frac{1}{720}\left[\frac{\omega^3}{m}\Bigl(153C_1-56C_2\Bigr)C_2+18g^2\right]r+O(r^2),\nonumber
\end{eqnarray}
where $r=t-t_0$, $\omega=\frac{1}{2m}g^2[3x_0^2-v^2]$, $C_2=C_1+\frac{g^2}{\omega^2}x_0$ and $C_3$ is a constant.
The expression for $A(t)$ can be derived from eq.~(\ref{well-hj-sol-co-1-sup-1}) straightforwardly
\begin{eqnarray}
\label{well-hj-sol-co-1-sim-sol}
A(t)=-\frac{1}{2}g^2(x_0^2-v^2)r+O(r^2).
\end{eqnarray}

\subsubsection{Derivation of the Lattice Lagrangian}\label{sec:2.2.2}

We have obtained a series solution for the Hamilton-Jacobi equation in section \ref{sec:2.2.1}. Then following the limiting procedures in \ref{sec:2.1.4},
we can derive the lattice Lagrangian. Because the additive properties of the limiting, we can implement the limiting procedures term by term. The results
are given by
%\begin{widetext}
\begin{eqnarray}
\label{well-latt-0}
\lim_{\substack{x_0{\rightarrow}x_a\\{t_0}{\rightarrow}t_a}}
\frac{S^{\ast}(x_b,t_b)-S^{\ast}(x_a,t_a)}{t_b-t_a}
=-\frac{1}{8}g^2\left(x_a^2-v^2\right)^2,\hspace{0mm}\\
\label{well-latt-1}
\lim_{\substack{x_0{\rightarrow}x_a\\{t_0}{\rightarrow}t_a}}
\frac{1}{t_b-t_a}\left[A(t_b)\left(x_b-f(t_b)\right)-A(t_a)\left(x_a-f(t_a)\right)\right]=-\frac{1}{2}g^2(x_a^2-v^2)(x_b-x_a),\hspace{0mm}\\
\label{well-latt-2}
\lim_{\substack{x_0{\rightarrow}x_a\\{t_0}{\rightarrow}t_a}}
\frac{1}{t_b-t_a}\left[R(t_b)
\left(x_b-f(t_b)\right)^2-R(t_a)\left(x_a-f(t_a)\right)^2\right]
=\frac{m}{2}\frac{(x_b-x_a)^2}{(t_b-t_a)^2}-\frac{g^2}{12}\left(3x_a^2-v^2\right)(x_b-x_a)^2,\hspace{0mm}\\
\label{well-latt-3}
\lim_{\substack{x_0{\rightarrow}x_a\\{t_0}{\rightarrow}t_a}}
\left[K(t_b)\left(x_b-f(t_b)\right)^3-K(t_a)\left(x_a-f(t_a)\right)^3\right]=-\frac{1}{24}g^2x_a(x_b-x_a)^3,\hspace{0mm}\\
\label{well-latt-4}
\lim_{\substack{x_0{\rightarrow}x_a\\{t_0}{\rightarrow}t_a}}
\frac{1}{t_b-t_a}\left[M(t_b)\left(x_b-f(t_b)\right)^4-M(t_a)\left(x_a-f(t_a)\right)^4\right]=-\frac{1}{40}g^2(x_b-x_a)^4.\hspace{0mm}
\end{eqnarray}
%\end{widetext}
Here we caution that we take the limit $x_0{\rightarrow}x_a$ at first, then take the limit $t_0{\rightarrow}t_a$. In the limiting procedures above,
we have set the constants of integral $C_2$ and $C_3$ to be zeros, so $C_1=-\frac{g^2}{\omega^2}x_0$, in order that the limits can be well defined;
Otherwise the limits will be singular function of $(t_0-t_a)$. The lattice Lagrangian is then found to be
\begin{eqnarray}
\label{well-latt-lag-hat}
\hat{L}_{\mathrm{lattice}}&=&-\frac{g^2}{8}\left[\Bigl(\frac{x_b+x_a}{2}-\frac{x_b-x_a}{2}\Bigr)^2-v^2\right]^2
-\frac{g^2}{2}(x_a^2-v^2)(x_b-x_a)+\frac{m}{2}\frac{(x_b-x_a)^2}{(t_b-t_a)^2}\\
&-&\frac{g^2}{2}x_a(x_b-x_a)^3-\frac{g^2}{40}(x_b-x_a)^4+\cdots.\nonumber
\end{eqnarray}
Again for infinitesimal lattice spacing $t_b-t_a=\epsilon$, eq.~(\ref{well-latt-lag-hat}) can be further approximated as
\begin{eqnarray}
\label{well-latt-lag}
\hat{L}_{\mathrm{lattice}}\xrightarrow[x_b\rightarrow x_a]{t_b{\rightarrow}t_a}\frac{m}{2}\frac{(x_b-x_a)^2}{(t_b-t_a)^2}-\frac{g^2}{8}
\left[\Bigl(\frac{x_b+x_a}{2}\Bigr)^2-v^2\right]^2,
\end{eqnarray}
which is the expected lattice version for the double-well Lagrangian (\ref{well-lag}). We saw that the terms of power higher than 3 disappear in the limit
$x_b\rightarrow x_a$. For the double-well potential, we can verify this phenomenon term by term. Thereafter, when we
find series solutions for nonlinear theories, we always suppose that this phenomenon holds. It seems to be plausible despite being difficult to prove.

\section{Examples of Scalar Field Theories}\label{sec:3}

In this section, we discuss scalar field theories. We try to apply the discussions of section~\ref{sec:2} to field theories. To make these applications feasible,
some new tools are required. These new tools are the covariant Hamilton theories or the De Donder-Weyl theories for field theories, based
on the multi-parameter generalization of Hilbert's independent integral. Taking scalar field theories for example, we will give a pedagogical introductions
of the De Donder-Weyl theory. For more details, see \cite{Courant:1962methods,Weyl:1934xb,weyl1935geodesic,Rund:1935,Kastrup:1982qq}. Based on these covariant Hamilton-Jacobi equations, we suggest an approach to derive discretized Lagrangians for field theories, similar to the procedures in section~\ref{sec:2}.

\subsection{Linear Theories: Massive Scalar Theories}\label{sec:3.1}

\subsubsection{Covariant Hamilton-Jacobi Equations}\label{sec:3.1.1}

For a massive free scalar field in four dimensional Minkowski space-time, its Lagrangian is given by
\begin{eqnarray}
\label{free-scalar-lag}
\mathscr{L}=\frac{1}{2}\eta^{\mu\nu}\partial_{\mu}\phi(x)\partial_{\nu}\phi(x)-\frac{1}{2} m^2 \phi(x)\phi(x).
\end{eqnarray}
Here and hereafter we use the Lorentzian metric $\eta=\mathrm{diag}(1,-1,-1,-1)$. Following De Donder and Weyl, the Legendre transformation is
performed with the manifest Lorentz covariance
\begin{eqnarray}
\label{free-scalar-legendre}
\pi_{\mu}=\frac{\partial\mathscr{L} }{\partial(\partial^{\mu}\phi)}=\partial_{\mu}\phi(x),
\end{eqnarray}
which is different from the Legendre transformation only performed in the time derivative of a field; While $\pi_{\mu}$ can be regarded as the covariant conjugate
momentum. Then the corresponding covariant Hamiltonian is found to be
\begin{eqnarray}
\label{free-scalar-ham}
\mathscr{H}=\pi_{\mu}\partial^{\mu}\phi(x)-\mathscr{L}=\frac{1}{2}\pi_{\mu}\pi^{\mu}+\frac{1}{2} m^2 \phi(x)\phi(x).
\end{eqnarray}
From this covariant Hamiltonian, we can get the Hamiltonian canonical equations
\begin{eqnarray}
\label{free-scalar-ham-canon-x}
\partial^{\mu}\phi(x)&=&\frac{\partial\mathscr{H} }{\partial\pi_{\mu}}=\pi^{\mu},\\
\label{free-scalar-ham-canon-p}
\partial^{\mu}\pi_{\mu}&=&-\frac{\partial\mathscr{H} }{\partial\phi}=-m^2\phi.
\end{eqnarray}
Obviously these canonical equations imply the following Euler-Lagrangian equation
\begin{eqnarray}
\label{free-scalar-euler}
\partial^{\mu}\partial_{\mu}\phi+m^2\phi=0,
\end{eqnarray}
as we expected.

The covariant Hamilton-Jacobi equation or the De Donder-Weyl equation can de derived as follows. Supposing the Lagrangian $\mathscr{L}$ is an independent integral
of Hilbert, that is, $\mathscr{L}$ can be expressed as the total derivative of a vector $S^{\mu}$
\begin{eqnarray}
\label{free-scalar-hilbert}
\mathscr{L}=\frac{dS^{\mu}}{dx^{\mu}}=\frac{\partial S^{\mu}}{\partial x^{\mu}}+\partial_{\mu}\phi(x)\frac{\partial S^{\mu}}{\partial \phi},
\end{eqnarray}
which also equals to
\begin{eqnarray}
\label{free-scalar-hilbert-mod}
\frac{\partial S^{\mu}}{\partial x^{\mu}}+\partial_{\mu}\phi(x)\frac{\partial S^{\mu}}{\partial \phi}-\mathscr{L}=0.
\end{eqnarray}
Designating
\begin{eqnarray}
\label{free-scalar-jacobi-mom}
\pi^{\mu}=\frac{\partial S^{\mu}}{\partial \phi},
\end{eqnarray}
we then get the covariant Hamilton-Jacobi equation
\begin{eqnarray}
\label{free-scalar-hj}
\frac{\partial S^{\mu}}{\partial x^{\mu}}+\mathscr{H}\left(\phi(x),\pi^{\mu}=\frac{\partial S^{\mu}}{\partial \phi}\right)=0.
\end{eqnarray}
As in classical mechanics, if we can find a complete solution for $S^{\mu}$ from the Hamilton-Jacobi equation (\ref{free-scalar-hj}),
then we can get a solution for the Euler-Lagrange equation (\ref{free-scalar-euler}). However, there are some differences between field theories and
classical mechanics. The reasons are that a classical mechanical system only depends on a single evolution parameter, the temporal variable $t$; While a system of
fields depend on the variables of time and space dimensions. Therefore, there are some integrability conditions involved for field theories. These
discussions can be understood as follows. If we find a solution for $S^{\mu}$, then we can determine $\pi^{\mu}$ by eq.~(\ref{free-scalar-jacobi-mom});
While on the other hand, $\pi^{\mu}$ is a conjugate momentum determined by the Legendre transformation (\ref{free-scalar-legendre}), so it can be expressed as
the total derivative of fields. Therefore $\pi^{\mu}$ determined by eq.~(\ref{free-scalar-jacobi-mom}) is connected to the total derivative of fields.
The foregoing are the reasons why the integrability conditions appear. For the free massive scalar field in this section, the integrability conditions are
straightforwardly found to be
\begin{eqnarray}
\label{free-scalar-jacobi-int}
\frac{d \pi_{\mu}}{d x^{\nu}}=\frac{d \pi_{\nu}}{d x^{\mu}},
\end{eqnarray}
or more specifically
\begin{eqnarray}
\label{free-scalar-jacobi-int-spe}
\frac{\partial \pi_{\mu}}{\partial x^{\nu}}+\partial_{\nu}\phi(x)\frac{\partial \pi_{\mu}}{\partial \phi}
=\frac{\partial \pi_{\nu}}{\partial x^{\mu}}+\partial_{\mu}\phi(x)\frac{\partial \pi_{\nu}}{\partial \phi}.
\end{eqnarray}
Here we noticed that $\pi^{\mu}$ is determined by eq.~(\ref{free-scalar-jacobi-mom}) but not eq.~(\ref{free-scalar-legendre}),
that is, $\pi^{\mu}$ is derived from the functional $S^{\mu}$. So these integrability conditions are basically some restriction conditions on $S^{\mu}$.
For more detailed discussions on integrability conditions, see \cite{weyl1935geodesic,vonRieth:1982ac,Kastrup:1982qq}.

\subsubsection{Solutions of Covariant Hamilton-Jacobi Equations}\label{sec:3.1.2}

For the Lagrangian (\ref{free-scalar-lag}) of a massive free scalar field, its De Donder-Weyl equation is given by
\begin{eqnarray}
\label{free-scalar-de-weyl}
\frac{\partial S^{\mu}}{\partial x^{\mu}}+\frac{1}{2}\frac{\partial S^{\mu}}{\partial \phi}\frac{\partial S_{\mu}}{\partial \phi}+\frac{1}{2} m^2 \phi(x)\phi(x)=0.
\end{eqnarray}
We employ the \lq\lq{embedding} method\rq\rq~introduced in section \ref{sec:2.2.1} to find its solution. In order to do that,
we need a solution of the Euler-Lagrange equation (\ref{free-scalar-euler}). We can find two types of solutions of equation (\ref{free-scalar-euler}):
\begin{eqnarray}
\label{free-scalar-euler-sol-1}
\mathrm{I}:\hspace{3mm}\varphi(x)&=&\varphi(z)\cos\Bigl(\frac{m}{\sqrt{\lambda}}r\Bigr),
\hspace{2mm}\lambda=k_{\mu}k^{\mu},\hspace{2mm}r=k_{\mu}(x-z)^{\mu},\\
\label{free-scalar-euler-sol-2}
\mathrm{II}:\hspace{2.5mm}\tilde{\varphi}(x)&=&\tilde{\varphi}(z)\frac{2}{m\sqrt{\xi}}\mathrm{BeeeslJ}(1,m\sqrt{\xi}),
\hspace{2mm}\xi=(x-z)_{\mu}(x-z)^{\mu},
\end{eqnarray}
where $k_{\mu}$ and $z_{\mu}$ are constant vectors and we have used the initial conditions to normalize the solutions. These two types of
solutions can both be used in the \lq\lq{embedding} method\rq\rq. We take the type (I) solution for example. Because eq.~(\ref{free-scalar-de-weyl}) only
contains quadratical terms, we can suppose that $S^{\mu}$ is given by
\begin{eqnarray}
\label{free-scalar-de-weyl-supp}
S^{\mu}&=&f^{\mu}(x)\Bigl[\phi(x)-\varphi(x)\Bigr]^2+h^{\mu}(x)\Bigl[\phi(x)-\varphi(x)\Bigr]+S^{\ast\mu}(x).
\end{eqnarray}
Substituting this expression into the De Donder-Weyl equation (\ref{free-scalar-de-weyl}), we can get a polynomial of $[\phi(x)-\varphi(x)]$. Letting the
coefficients of this polynomial to be zero, we get
%\begin{widetext}
\begin{eqnarray}
\label{free-scalar-de-weyl-supp-co-0}
\Bigl[\phi(x)-\varphi(x)\Bigr]^0:\hspace{4mm}\partial_{\mu} S^{\ast\mu}-\partial_{\mu}\varphi(x)h^{\mu}+\frac{1}{2}h_{\mu}h^{\mu}+\frac{1}{2}m^2\varphi^2(x)&=&0,\\
\label{free-scalar-de-weyl-supp-co-1}
\Bigl[\phi(x)-\varphi(x)\Bigr]^1:\hspace{10mm}\partial_{\mu} h^{\mu}+2f^{\mu}[-\partial_{\mu}\varphi(x)+h_{\mu}]+m^2\varphi(x)&=&0,\\
\label{free-scalar-de-weyl-supp-co-2}
\Bigl[\phi(x)-\varphi(x)\Bigr]^2:\hspace{34.5mm}\partial_{\mu} f^{\mu}+2f_{\mu}f^{\mu}+\frac{1}{2}m^2&=&0.
\end{eqnarray}
%\end{widetext}
Similar to that in section \ref{sec:2.2.1}, we can suppose
\begin{eqnarray}
\label{free-scalar-de-weyl-supp-co-x}
h_{\mu}&=&\partial_{\mu}\varphi(x),\\
\label{free-scalar-de-weyl-supp-co-p}
\frac{d h^{\mu}}{d x^{\mu}}&=&-m^2\varphi(x),
\end{eqnarray}
then eq.~(\ref{free-scalar-de-weyl-supp-co-1}) gets satisfied. Obviously, Eqs.~(\ref{free-scalar-de-weyl-supp-co-x}) and (\ref{free-scalar-de-weyl-supp-co-p}) are
just the Hamiltonian canonical equations (\ref{free-scalar-ham-canon-x}) and (\ref{free-scalar-ham-canon-p}). They are consistent because $\varphi(x)$ is a solution
of the Euler-Lagrange equation. While eq.~(\ref{free-scalar-de-weyl-supp-co-0}) is
transformed to be
\begin{eqnarray}
\label{free-scalar-de-weyl-supp-co-0-trans}
\frac{d S^{\ast\mu}}{d x^{\mu}}&=&\partial_{\mu}\varphi(x)h^{\mu}-\frac{1}{2}h_{\mu}h^{\mu}-\frac{1}{2}m^2\varphi^2(x)
={\mathscr{L}}^{\ast}\Bigl(\varphi(x)\Bigr),
\end{eqnarray}
whose right-hand side is the Lagrangian evaluated at $\varphi(x)$. The solutions for $f^{\mu}$, $h^{\mu}$ and $S^{\ast\mu}$ can be given by
\begin{eqnarray}
\label{free-scalar-de-weyl-sol-f}
f^{\mu}(x)&=&\frac{1}{2}\frac{m}{\sqrt{\lambda}}\cot\Bigl(\frac{m}{\sqrt{\lambda}}r\Bigr)k^{\mu},\\
\label{free-scalar-de-weyl-sol-h}
h^{\mu}(x)&=&-\frac{m}{\sqrt{\lambda}}\sin\Bigl(\frac{m}{\sqrt{\lambda}}r\Bigr)k^{\mu},\\
\label{free-scalar-de-weyl-sol-s}
S^{\ast\mu}(x)&=&-\frac{m}{4\sqrt{\lambda}}\sin\Bigl(\frac{2m}{\sqrt{\lambda}}r\Bigr)\varphi^2(z)k^{\mu},
\end{eqnarray}
where $r$ has been defined by eq.~(\ref{free-scalar-euler-sol-1}).

\subsubsection{Derivation of the Discretized Lagrangian}\label{sec:3.1.3}

In order to derive the discretized Lagrangian for field theories, we need a formula similar to eq.~(\ref{dis-lag-sol-hj-def}). However, some differences emerge
in field theories, because what we have in field theories is the vector functional $S^{\mu}$. So we need a multi-dimensional exploration for the formula
(\ref{dis-lag-sol-hj-def}). Inspired by the designation of the independent integral of Hilbert, that is, eq.~(\ref{free-scalar-hilbert}), we suggest the following
definition for discretized Lagrangian
\begin{eqnarray}
\label{dis-field-sol-hj-def}
\tilde{\mathscr{L}}_{\mathrm{lattice}}&=&\sum_{i=0}^3 \frac{S^{i}(y^{i})-S^{i}(x^{i})}{y^{i}-x^{i}},
\end{eqnarray}
where
\begin{eqnarray}
\label{dis-field-sol-hj-def-total}
\frac{S^{1}(y^{1})-S^{1}(x^{1})}{y^{1}-x^{1}}
=\frac{1}{y^{1}-x^{1}}\Bigl[S^{1}\left(x^{0},y^{1},x^{2},x^{3},\phi(x^{0},y^{1},x^{2},x^{3})\right)
-S^{1}\left(x^{0},x^{1},x^{2},x^{3},\phi(x^{0},x^{1},x^{2},x^{3})\right)\Bigr].\nonumber\\
\end{eqnarray}
The definitions for other indices follow similarly. Obviously, in the limit of $y^{i}-x^{i}\rightarrow 0$, this definition is just the independent
integral of Hilbert (\ref{free-scalar-hilbert}).
For the solutions in eqs.~(\ref{free-scalar-de-weyl-sol-f}), (\ref{free-scalar-de-weyl-sol-h}) and (\ref{free-scalar-de-weyl-sol-s}), employing the definition in (\ref{dis-field-sol-hj-def}), then taking the limits of $z^{i}\rightarrow x^{i} $ and $\varphi(z)\rightarrow\phi(x)$, we can get
\begin{eqnarray}
\label{dis-field-sol-hj-def-limit}
\hat{\mathscr{L}}_{\mathrm{lattice}}&=&\lim_{z^i\rightarrow x^i}\lim_{\varphi(z^i)\rightarrow\phi(x^i)}\tilde{\mathscr{L}}_{\mathrm{lattice}}\\
&=&\sum_{i=0}^3 \frac{1}{2}\frac{1}{\epsilon}\frac{m}{\sqrt{\lambda}}\cot\Bigl(\frac{m}{\sqrt{\lambda}}\epsilon k_{i} \Bigr)k^{i}
\Bigl[\phi(y^{i})-\phi(x^{i})\Bigr]^{2}\nonumber\\
&-&\sum_{i=0}^3 \frac{1}{\epsilon}\frac{m}{\sqrt{\lambda}}\sin\Bigl(\frac{m}{\sqrt{\lambda}}\epsilon k_{i} \Bigr)k^{i}\Bigl[\phi(y^{i})-\phi(x^{i})\Bigr]
-\sum_{i=0}^3\frac{1}{\epsilon}\frac{m}{4\sqrt{\lambda}}\phi^2(x^i)\sin\Bigl(\frac{2m}{\sqrt{\lambda}}\epsilon k_{i}\Bigr)k^{i}.\nonumber
\end{eqnarray}
Here we have supposed the symmetrical lattice spacing, that is, $y^{i}-x^{i}=\epsilon$. Furthermore,
supposing infinitesimal lattice spacing $\epsilon\rightarrow 0$
and $\phi(y^i)-\phi(x^i)\rightarrow 0$, we get the final version of the discretized Lagrangian
\begin{eqnarray}
\label{dis-field-sol-hj-def-final}
\mathscr{L}_{\mathrm{lattice}}=\lim_{y^i\rightarrow x^i}\lim_{\phi(y^i)\rightarrow\phi(x^i)}\hat{\mathscr{L}}_{\mathrm{lattice}}
=\sum_{i=0}^3 {\mathrm{sgn}(i)}\frac{1}{2}\frac{1}{\epsilon^2}\Bigl[\phi(y^{i})-\phi(x^{i})\Bigr]^{2}
-\frac{1}{2}m^2\Bigl[\frac{\phi(y)+\phi(x)}{2}\Bigr]^2.\nonumber
\end{eqnarray}
Here we have defined the function
\begin{eqnarray}
\label{sgn-def}
\mathrm{sgn}(i)
=\begin{cases}
1,~~~&i=0, \\
-1,~~~&i=1,2,3.
\end{cases}
\end{eqnarray}
Obviously, eq.~(\ref{dis-field-sol-hj-def-final}) is the lattice version of the Lagrangian (\ref{free-scalar-lag}) as we expected.
The appearance of the $\mathrm{sgn}(i)$ is because we work on Minkowski space-time with a Lorentzian metric.
For infinitesimal $\epsilon\rightarrow 0$, we can make the replacement
\begin{eqnarray}
\label{dis-field-sol-hj-def-final-replace}
\frac{\phi(y^{i})-\phi(x^{i})}{\epsilon}\xrightarrow[]{\epsilon\rightarrow 0}\partial_{i}\phi(x),
\end{eqnarray}
then the lattice Lagrangian will approximate to the continuous one
\begin{eqnarray}
\label{dis-field-sol-hj-def-final-continue}
\mathscr{L}_{\mathrm{lattice}}\xrightarrow[]{\epsilon \rightarrow 0}
\frac{1}{2}\eta^{\mu\nu}\partial_{\mu}\phi(x)\partial_{\nu}\phi(x)-\frac{1}{2} m^2 \phi(x)\phi(x),\hspace{2mm}
\end{eqnarray}
which is of course the Lagrangian (\ref{free-scalar-lag}) we began with.

Moreover, we should mention that the Lagrangian (\ref{dis-field-sol-hj-def-final}) is not the only lattice Lagrangian we can obtain by following
the foregoing procedure. The reason is that we can obtain more solutions besides the solutions in equations (\ref{free-scalar-de-weyl-sol-f}),
(\ref{free-scalar-de-weyl-sol-h}) and (\ref{free-scalar-de-weyl-sol-s}). Actually, we can get two more solutions of
eq.~(\ref{free-scalar-de-weyl-supp-co-2}) for $f^{\mu}$
\begin{eqnarray}
\label{free-scalar-de-weyl-sol-f-1}
f_1^{\mu}(x)&=&-\frac{1}{2}\frac{m}{\sqrt{\lambda}}\tanh\Bigl(\frac{m}{\sqrt{\lambda}}r\Bigr)k^{\mu},\\
\label{free-scalar-de-weyl-sol-f-2}
f_2^{\mu}(x)&=&-\frac{1}{2}\frac{m}{\sqrt{\xi}}\frac{\mathrm{BesselY}(2,m\sqrt{\xi})}{\mathrm{BesselY}(1,m\sqrt{\xi})}(x-z)^{\mu},
\end{eqnarray}
where $r$ and $\xi$ have been defined in eqs.~(\ref{free-scalar-euler-sol-1}) and (\ref{free-scalar-euler-sol-2}).
Associated with solutions (\ref{free-scalar-de-weyl-sol-h}) and (\ref{free-scalar-de-weyl-sol-s}), eqs.~(\ref{free-scalar-de-weyl-sol-f-1})
and (\ref{free-scalar-de-weyl-sol-f-2}) both construct new solutions for $S^{\mu}$. Following the limiting procedures as we just did above,
we can derive two new Lagrangians from these two new solutions. They are given by
\begin{eqnarray}
\label{dis-field-sol-hj-def-final-continue-1}
f_1^{\mu}:\hspace{1mm}\mathscr{L}_{1}&=&-\frac{1}{2} m^2 \phi(x)\phi(x),\\
\label{dis-field-sol-hj-def-final-continue-2}
f_2^{\mu}:\hspace{1mm} \mathscr{L}_{2}&=&-\frac{1}{2}\eta^{\mu\nu}\partial_{\mu}\phi(x)\partial_{\nu}\phi(x)-\frac{1}{2} m^2 \phi(x)\phi(x).\hspace{4mm}
\end{eqnarray}
The first new one has no kinetic term because $f_1^{\mu}$ does not contribute in the limit of $\epsilon\rightarrow 0$; While the second new one has
the minus kinetic term, that is, a ghost kinetic term because $f_2^{\mu}$ yields the minus kinetic term. We might expect the integrability condition introduced
in (\ref{free-scalar-jacobi-int}) to kill these two new solutions, but they both satisfy the integrability condition. We will see that
this phenomenon happens for linear theories universally. So far we can not figure out they could correspond to two new sectors of quantum theories
we can derive from solutions of the De Donder-Weyl equations, or they mean the procedure we have employed is incomplete
so we need some more criteria to select the physical solution.

\subsection{Nonlinear Theories: Scalar Theories with $\lambda \phi^4 $ Potential}\label{sec:3.2}

In this section, we discuss nonlinear field theories. These discussions are in conjunction with that of linear theories. The differences are we can get exact solutions
for linear theories, but we can only get series solutions for nonlinear theories.

\subsubsection{De Donder-Weyl Equation and Its Solution}\label{sec:3.2.1}

We consider a scalar field theory of $\lambda \phi^4 $ potential in four dimensional Minkowski space-time, its Lagrangian is given by
\begin{eqnarray}
\label{non-scalar-lag}
\mathscr{L}=\frac{1}{2}\eta^{\mu\nu}\partial_{\mu}\phi(x)\partial_{\nu}\phi(x)
-\frac{1}{2} m^2 \phi(x)\phi(x)-\frac{\lambda}{4!} \phi^4(x)-\Lambda,
\end{eqnarray}
where we include a density of vacuum energy $\Lambda$ for generality. Performing the Legendre transformation
\begin{eqnarray}
\label{non-scalar-legendre}
\pi_{\mu}=\frac{\partial\mathscr{L}}{\partial(\partial^{\mu}\phi)}=\partial_{\mu}\phi(x),
\end{eqnarray}
we get the covariant Hamiltonian
\begin{eqnarray}
\label{non-scalar-ham}
\mathscr{H}=\frac{1}{2}\pi_{\mu}\pi^{\mu}+\frac{1}{2} m^2 \phi(x)\phi(x)+\frac{\lambda}{4!} \phi^4(x)+\Lambda.
\end{eqnarray}
The corresponding Hamiltonian canonical equations are
\begin{eqnarray}
\label{non-scalar-ham-canon-x}
\partial^{\mu}\phi(x)&=&\frac{\partial\mathscr{H} }{\partial\pi_{\mu}}=\pi^{\mu},\\
\label{non-scalar-ham-canon-p}
\partial^{\mu}\pi_{\mu}&=&-\frac{\partial\mathscr{H} }{\partial\phi}=-m^2\phi-\frac{\lambda}{3!} \phi^3(x).
\end{eqnarray}
They imply the following Euler-Lagrangian equation
\begin{eqnarray}
\label{non-scalar-euler}
\partial^{\mu}\partial_{\mu}\phi+m^2\phi+\frac{\lambda}{3!} \phi^3(x)=0.
\end{eqnarray}
In eq.~(\ref{non-scalar-ham}), designating
\begin{eqnarray}
\label{non-scalar-hj-des}
\pi_{\mu}=\frac{\partial S_{\mu}}{\partial\phi},
\end{eqnarray}
we get its De Donder-Weyl equation
\begin{eqnarray}
\label{non-scalar-weyl}
 \partial_{\mu}S^{\mu}+\frac{1}{2}\frac{\partial S_{\mu} }{\partial\phi}\frac{\partial S^{\mu} }{\partial\phi}&+&\frac{1}{2} m^2 \phi(x)\phi(x)
 +\frac{\lambda}{4!} \phi^4(x)+\Lambda=0.
\end{eqnarray}

Following the procedures in section \ref{sec:3.1.2}, we employ the \lq\lq{embedding} method\rq\rq~to find the solution of eq.~(\ref{non-scalar-weyl}). As in
section \ref{sec:3.1.2}, we need a solution of the Euler-Lagrange equation, which can be given by
\begin{eqnarray}
\label{non-scalar-euler-sol-1}
\varphi(x)=\varphi(z)\mathrm{JacobiDN}\left(\frac{\sqrt{\lambda}\varphi(z)}{2\sqrt{3\sigma}}r,k\right), \hspace{2mm}
k=\sqrt{2+\frac{12m^2}{\lambda\varphi^2(z)}},\hspace{2mm}r=p_{\mu}(x-z)^{\mu},\hspace{2mm}\sigma=p_{\mu}p^{\mu}.
\end{eqnarray}
Here $p_{\mu}$ and $z_{\mu}$ are constant vectors, and we have used  $\varphi(x)\vert_z=\varphi(z)$ to normalize the solution.
According to the \lq\lq{embedding} method\rq\rq, we suppose a series solution for $S_{\mu}$
\begin{eqnarray}
\label{non-scalar-de-weyl-supp}
S^{\mu}&=&S^{\ast\mu}(x)+P^{\mu}(x)\left[\phi(x)-\varphi(x)\right]+R^{\mu}(x)\left[\phi(x)-\varphi(x)\right]^2\\
&+&K^{\mu}(x)\left[\phi(x)-\varphi(x)\right]^3
+M^{\mu}(x)\left[\phi(x)-\varphi(x)\right]^4+N^{\mu}(x)\left[\phi(x)-\varphi(x)\right]^5+\cdots.\nonumber
\end{eqnarray}
Substituting this expression into eq.~(\ref{non-scalar-weyl}), we get a series expression of $\left[\phi(x)-\varphi(x)\right]$.
Supposing the coefficients of this series to be zeros term by term, we get
%\begin{widetext}
\begin{eqnarray}
\label{non-scalar-de-weyl-supp-co-0}
\left[\phi(x)-\varphi(x)\right]^0:\hspace{0mm}\partial_{\mu} S^{\ast\mu}-\partial_{\mu}\varphi(x)P^{\mu}+\frac{1}{2}P_{\mu}P^{\mu}+\frac{1}{2}m^2\varphi^2(x)+
\frac{\lambda}{4!} \varphi^4(x)+\Lambda&=&0,\\
\label{non-scalar-de-weyl-supp-co-1}
\left[\phi(x)-\varphi(x)\right]^1:\hspace{13.5mm}\partial_{\mu} P^{\mu}+2R^{\mu}[-\partial_{\mu}\varphi(x)+P_{\mu}]+m^2\varphi(x)+\frac{\lambda}{6} \varphi^3(x)&=&0,\\
\label{non-scalar-de-weyl-supp-co-2}
\left[\phi(x)-\varphi(x)\right]^2:\hspace{2mm}\partial_{\mu} R^{\mu}+3K^{\mu}[-\partial_{\mu}\varphi(x)+P_{\mu}]+2R_{\mu}R^{\mu}+\frac{1}{2}m^2
+\frac{\lambda}{4} \varphi^2(x)&=&0,\\
\label{non-scalar-de-weyl-supp-co-3}
\left[\phi(x)-\varphi(x)\right]^3:\hspace{13.5mm}\partial_{\mu} K^{\mu}+4M^{\mu}[-\partial_{\mu}\varphi(x)+P_{\mu}]+6R_{\mu}K^{\mu}
+\frac{\lambda }{6}\varphi(x)&=&0,\\
\label{non-scalar-de-weyl-supp-co-4}
\left[\phi(x)-\varphi(x)\right]^4:\hspace{2mm}\partial_{\mu} M^{\mu}+5N^{\mu}[-\partial_{\mu}\varphi(x)+P_{\mu}]+\frac{9}{2}K_{\mu}K^{\mu}
+8M_{\mu}R^{\mu}+\frac{\lambda}{24}&=&0,
\end{eqnarray}
%\end{widetext}
where terms of power higher than 4 are omitted for convenience. By supposing the self-consistent canonical equations
\begin{eqnarray}
\label{non-scalar-ham-canon-sol-x}
\partial^{\mu}\varphi(x)&=&P^{\mu},\\
\label{non-scalar-ham-canon-sol-p}
\partial^{\mu}P_{\mu}&=&-m^2\varphi-\frac{\lambda}{3!}\varphi^3(x),
\end{eqnarray}
eq.~(\ref{non-scalar-de-weyl-supp-co-1}) gets satisfied, and eq.~(\ref{non-scalar-de-weyl-supp-co-0}) is transformed to be
\begin{eqnarray}
\label{non-scalar-de-weyl-supp-co-0-tran}
\partial_{\mu} S^{\ast\mu}=\partial_{\mu}\varphi(x)P^{\mu}-\frac{1}{2}P_{\mu}P^{\mu}-\frac{1}{2}m^2\varphi^2(x)
-\frac{\lambda}{4!} \varphi^4(x)-\Lambda=\mathscr{L}^{\ast}\Bigl(\varphi(x)\Bigr).
\end{eqnarray}
Eqs.~(\ref{non-scalar-de-weyl-supp-co-2}), (\ref{non-scalar-de-weyl-supp-co-3}) and (\ref{non-scalar-de-weyl-supp-co-4}) are simplified to be
\begin{eqnarray}
\label{non-scalar-de-weyl-supp-co-2-sim}
\partial_{\mu} R^{\mu}+2R_{\mu}R^{\mu}+\frac{1}{2}m^2+\frac{\lambda}{4} \varphi^2(x)&=&0,\\
\label{non-scalar-de-weyl-supp-co-3-sim}
\partial_{\mu} K^{\mu}+6R_{\mu}K^{\mu}+\frac{\lambda }{6}\varphi(x)&=&0,\\
\label{non-scalar-de-weyl-supp-co-4-sim}
\partial_{\mu} M^{\mu}+\frac{9}{2}K_{\mu}K^{\mu}+8M_{\mu}R^{\mu}+\frac{\lambda}{24}&=&0.
\end{eqnarray}
The exact solutions of these equations are difficult to derive. However, the behavior of solutions around small $r$ is enough for our purpose. For small $r$, we can replace $\varphi(x)$ with $\varphi(z)$ in eqs.~(\ref{non-scalar-de-weyl-supp-co-2-sim}), (\ref{non-scalar-de-weyl-supp-co-3-sim})
and (\ref{non-scalar-de-weyl-supp-co-4-sim}). Then we can get the following series solutions for $R^{\mu}$, $K^{\mu}$ and $M^{\mu}$
\begin{eqnarray}
\label{non-scalar-de-weyl-sol-2}
R^{\mu}&=&\left[\frac{1}{2}\frac{1}{r}-\frac{1}{6}{\omega}r\right] p^{\mu}+O(r^3),\\
\label{non-scalar-de-weyl-sol-3}
K^{\mu}&=&\left[\frac{C_2}{r^3}+\frac{1}{2}\omega\frac{C_2}{r}+\frac{C_1}{8}\omega^2r\right] p^{\mu}+O(r^3),\\
\label{non-scalar-de-weyl-sol-4}
M^{\mu}&=&\left[\frac{9}{2}\frac{C^2_2}{r^5}+\frac{C_3}{r^4}+\frac{3\omega}{2}\frac{C^2_2}{r^3}+\frac{2\omega}{3}\frac{C_3}{r^2}\right] p^{\mu}\\
&+&\frac{\omega^2}{24}\Bigl(-9C_1+7C_2\Bigr)C_2\frac{1}{r}p^{\mu}+\frac{2\omega^2}{9}C_3p^{\mu}
-\left[\Bigl(\frac{17}{80}C_1-\frac{7}{90}C_2\Bigr)C_2+\frac{1}{120}\frac{\lambda}{\sigma}\right]r p^{\mu}+O(r^2),\nonumber
\end{eqnarray}
where $C_2= C_1+\frac{\lambda\varphi(z)}{3\omega^2\sigma}$, $\omega=\sqrt{\frac{m^2}{\sigma}+\frac{\lambda\varphi^2(z)}{2\sigma}}$ and $C_3$ is a constant. While the series expressions of $S^{\ast\mu}$ and $P^{\mu}$ can be derived from equations (\ref{non-scalar-ham-canon-sol-x}) and (\ref{non-scalar-de-weyl-supp-co-0-tran})
\begin{eqnarray}
\label{non-scalar-de-weyl-sol-0}
S^{\ast\mu}&=&\left[-\frac{m^2}{2}\varphi^2(z)
-\frac{\lambda}{4!} \varphi^4(z)-\Lambda\right]\frac{r}{\sigma} p^{\mu}+O(r^3),\hspace{3mm}\\
\label{non-scalar-de-weyl-sol-1}
P^{\mu}&=&\varphi(z)\left[-m^2-\frac{\lambda}{6}\varphi^2(z)\right]\frac{r}{\sigma}p^{\mu}+O(r^3),
\end{eqnarray}
where $\sigma=p^{\mu}p_{\mu}$. Here we notice that the foregoing solutions are very similar to that which we obtained in section \ref{sec:2.2.1}.

\subsubsection{Derivation of the Discretized Lagrangian}\label{sec:3.2.2}

Having obtained the solutions in the last subsection, it is straightforward to derive the discretized Lagrangian following the procedures
in section \ref{sec:3.1.3}. By the definition of eq.~(\ref{dis-field-sol-hj-def}), taking limits term by term as we did in section \ref{sec:2.2.2},
then we can get the lattice Lagrangian
\begin{eqnarray}
\label{latt-field-sol-hj-def-limit-non}
\hat{\mathscr{L}}_{\mathrm{lattice}}&=&\lim_{z^i\rightarrow x^i}\lim_{\varphi(z^i)\rightarrow\phi(x^i)}\tilde{\mathscr{L}}_{\mathrm{lattice}}
=\lim_{z^i\rightarrow x^i}\lim_{\varphi(z^i)\rightarrow\phi(x^i)} \sum_{i=0}^3 \frac{S^{i}(y^{i})-S^{i}(x^{i})}{y^{i}-x^{i}}\\
&=&\sum_{i=0}^3\frac{1}{\sigma}p_{i}p^{i}\left[-\frac{1}{2} m^2 \phi^2(x^{i})-\frac{\lambda}{4!} \phi^4(x^{i})-\Lambda\right]
-\sum_{i=0}^3\frac{1}{\sigma}p_{i}p^{i}\phi(x^{i})\left[m^2+\frac{\lambda}{6}\phi^2(x^{i})\right]\left[\phi(y^{i})-\phi(x^{i})\right]\nonumber\\
&+&\sum_{i=0}^3 \frac{1}{2}\frac{1}{\epsilon^2}\frac{p^{i}}{p_{i}}\left[\phi(y^{i})-\phi(x^{i})\right]^{2}
-\frac{1}{6}\sum_{i=0}^3 p_{i}p^{i}\left[\phi(y^{i})-\phi(x^{i})\right]^{2}\left[\frac{m^2}{\sigma}+\frac{\lambda\phi^2(x^{i})}{2\sigma}\right]\nonumber\\
&-&\frac{1}{24}\sum_{i=0}^3\frac{1}{\sigma} p_{i}p^{i}\phi(x^{i})\left[\phi(y^{i})-\phi(x^{i})\right]^{3}
-\frac{\lambda}{120}\sum_{i=0}^3\frac{1}{\sigma} p_{i}p^{i}\left[\phi(y^{i})-\phi(x^{i})\right]^{4}+\cdots.\nonumber
\end{eqnarray}
In derivations of this Lagrangian, we have supposed the symmetrical lattice spacing  $y^{i}-x^{i}=\epsilon$, and in the limiting procedures above,
we have set the constants of integral $C_2$ and $C_3$ to be zeros, so $C_1=-\frac{\lambda\varphi(z)}{3\omega^2\sigma}$, in order that the limits can be well defined. From this lattice Lagrangian, further assumptions on infinitesimal lattice spacing  $\epsilon$ will lead to the lattice Lagrangian of $\lambda \phi^4$ interactions as we expected.

\section{Topologically Massive Gauge Theory}\label{sec:4}

In this section, we discuss the topologically massive gauge theory introduced in \cite{Deser:1982vy,Deser:1981wh}. For an Abelian gauge field
in three dimensional Minkowski space-time supplemented with the Chern-Simons topological term, its Lagrangian is given by
\begin{eqnarray}
\label{gauge-lag}
\mathscr{L}&=&\frac{1}{4\Lambda}F_{\mu\nu}F^{\mu\nu}+\frac{\theta}{4}\varepsilon^{\mu\nu\lambda}F_{\mu\nu}A_{\lambda},\\
F_{\mu\nu}&=&\partial_{\mu}A_{\nu}-\partial_{\nu}A_{\mu},\nonumber
\end{eqnarray}
where the three dimensional Lorentzian metric is $\eta=\mathrm{diag}(1,-1,-1)$, and $\varepsilon^{\mu\nu\lambda}$ is the totally antisymmetrical tensor with the convention
$\varepsilon^{012}=1$. In three dimensions, $\Lambda$ is a minus constant of mass dimension, and $\theta$ is dimensionless. In \cite{Deser:1982vy,Deser:1981wh}, the authors
show that the gauge field excitations are massive. The massive excitations can be understood as follows. The Euler-Lagrange equation for (\ref{gauge-lag}) is
\begin{eqnarray}
\label{gauge-lag-euler}
\partial_{\mu}F^{\mu\nu}-\frac{\theta}{2}\Lambda\varepsilon^{\nu\alpha\beta}F_{\alpha\beta}=0.
\end{eqnarray}
The dual field strength of $F_{\alpha\beta}$ is defined as
\begin{eqnarray}
\label{gauge-field-dual-def}
{}^\ast{F^{\mu}}=\frac{1}{2}\varepsilon^{\mu\alpha\beta}F_{\alpha\beta},
\end{eqnarray}
then $F_{\alpha\beta}$ can also be expressed by its dual field strength
\begin{eqnarray}
\label{gauge-field-dual-def-dual}
F^{\mu\nu}=\varepsilon^{\mu\nu\alpha}{}^\ast{F_{\alpha}}.
\end{eqnarray}
Substituting this expression into the equation of motion (\ref{gauge-lag-euler}), we get an equation of motion for the dual field strength
\begin{eqnarray}
\label{gauge-lag-euler-dual}
\Bigl(\varepsilon^{\nu\mu\alpha}\partial_{\mu}+\theta\Lambda\eta^{\nu\alpha}\Bigr){}^\ast{F_{\alpha}}=0,
\end{eqnarray}
which furthermore implies
\begin{eqnarray}
\label{gauge-lag-euler-dual-oeder-2}
\Bigl(\varepsilon_{\beta\lambda\nu}\partial^{\lambda}-\theta\Lambda\eta_{\nu\beta}\Bigr)
\Bigl(\varepsilon^{\nu\mu\alpha}\partial_{\mu}+\theta\Lambda\eta^{\nu\alpha}\Bigr){}^\ast{F_{\alpha}}
&=&\biggl[\partial_{\alpha}\partial_{\beta}-\eta_{\alpha\beta}\Bigl(\partial_{\mu}\partial^{\mu}+\theta^2\Lambda^2\Bigr)\biggr]{}^\ast{F^{\alpha}}\nonumber\\
&=&-\Bigl(\partial_{\mu}\partial^{\mu}+\theta^2\Lambda^2\Bigr){}^\ast{F^{\beta}}=0.
\end{eqnarray}
In the calculation above, we have used the Bianchi identity
\begin{eqnarray}
\label{gauge-dual-bianchi}
\partial_{\alpha}{}^\ast{F^{\alpha}}=0.
\end{eqnarray}
Eq.~(\ref{gauge-lag-euler-dual-oeder-2}) clearly demonstrates that the dual field ${}^\ast{F^{\beta}}$ is massive. For another analysis based on the Dyson-Schwinger
equation, see \cite{Aguilar:2010zx}.

\subsection{De Donder-Weyl Equation for Gauge Theories}\label{sec:4.1}

We start with the Lagrangian (\ref{gauge-lag}) supplemented with the gauge fixed term
\begin{eqnarray}
\label{gauge-lag-gauge-fix}
\mathscr{L}=\frac{1}{4\Lambda}F_{\mu\nu}F^{\mu\nu}+\frac{\theta}{4}\varepsilon^{\mu\nu\lambda}F_{\mu\nu}A_{\lambda}
+\frac{1}{2\xi}\partial_{\nu}A_{\mu}\partial^{\mu}A^{\nu}.\hspace{4mm}
\end{eqnarray}
The reason for supplementing the gauge fixed term will be clear soon later. Performing the covariant Legendre transformation
\begin{eqnarray}
\label{gauge-legendre}
\pi^{\mu\nu}=\frac{\partial \mathscr{L}}{\partial(\partial_{\mu}A_{\nu})}
=\frac{1}{\Lambda}F^{\mu\nu}+\frac{1}{\xi}\partial^{\nu}A^{\mu}+\frac{\theta}{2}\varepsilon^{\mu\nu\lambda}A_{\lambda}.
\end{eqnarray}
Without the gauge fixed term, because $F^{\mu\nu}$ is a antisymmetrical tensor, we could get a constraint relation for the symmetric part of $\pi^{\mu\nu}$
\begin{eqnarray}
\label{gauge-legendre-restr}
\pi^{\mu\nu}+\pi^{\nu\mu}=0.
\end{eqnarray}
As is well known in conventional Hamiltonian theories, the appearance of the constraint condition implies the singularity of the Legendre transformation,
which is ascribed to the gauge invariance of the original theories. For the gauge fixed Lagrangian (\ref{gauge-lag-gauge-fix}), the constraint
condition (\ref{gauge-legendre-restr}) disappears\footnote{In this paper, we always add a gauge fixed term to cure the problem of irregularity of Lagrangian.
Another approach is adding Lagrange multipliers to the Hamiltonian to handle the gauge constraints as that did by
Dirac and Bergmann \cite{Dirac:1951zz,Dirac:1958sq,Bergmann:1949}.
More geometrical analysis on the constraints have been carried out in \cite{Gotay:2004ib}.}.
Here and hereafter we use the gauge $\xi=\Lambda$, so the derivative of gauge field can be expressed by
its conjugate momentum as follows
\begin{eqnarray}
\label{gauge-legendre-gauge-fix}
\partial^{\mu}A^{\nu}=\Lambda\pi^{\mu\nu}-\frac{\Lambda}{2}\theta\varepsilon^{\mu\nu\lambda}A_{\lambda}.
\end{eqnarray}
The covariant Hamiltonian is derived as
\begin{eqnarray}
\label{gauge-ham}
\mathscr{H}&=&\partial^{\mu}A^{\nu}\pi_{\mu\nu}-\mathscr{L}
=\frac{\Lambda}{2}\pi^{\mu\nu}\pi_{\mu\nu}
-\frac{\Lambda}{2}\theta\varepsilon^{\mu\nu\lambda}\pi_{\mu\nu}A_{\lambda}+\frac{\Lambda}{4}\theta^2A_{\mu}A^{\mu}.
\end{eqnarray}
The canonical Hamiltonian equation is given by
\begin{eqnarray}
\label{gauge-ham-cano-x}
\partial_{\mu}A_{\nu}&=&\frac{\partial\mathscr{H}}{\partial\pi^{\mu\nu}}=\Lambda\pi_{\mu\nu}-\frac{\Lambda}{2}\theta\varepsilon_{\mu\nu\lambda}A^{\lambda},\\
\label{gauge-ham-cano-p}
\partial_{\mu}\pi^{\mu\nu}&=&-\frac{\partial\mathscr{H}}{\partial A_{\nu}}=\frac{\Lambda}{2}\theta\varepsilon_{\alpha\beta\nu}\pi^{\alpha\beta}
-\frac{\Lambda}{2}\theta^2 A^{\nu},
\end{eqnarray}
which together means that the Euler-Lagrange equation is
\begin{eqnarray}
\label{gauge-ham-cano-euler}
\partial_{\mu}\partial^{\mu}A^{\nu}-\Lambda\theta\varepsilon^{\alpha\beta\nu}\partial_{\alpha}A_{\beta}=0,
\end{eqnarray}
which is the gauge fixed version of eq.~(\ref{gauge-lag-euler}). We noticed that a mass term appears in the covariant Hamiltonian (\ref{gauge-ham}), which
might be ascribed to the massive property of this gauge theory as we discussed in the beginning of section \ref{sec:4}.
However, the contribution of this mass term is canceled by that of the second topological term in (\ref{gauge-ham})
secretly. So the Euler-Lagrange equation is still of the same massless form as that of eq.~(\ref{gauge-lag-euler}).

Similar to the discussions in section \ref{sec:3.1.1}, we designate the Lagrangian to be the independent integral of Hilbert
\begin{eqnarray}
\label{gauge-hibert}
\mathscr{L}=\frac{d S^{\mu}}{d x^{\mu}}= \partial_{\mu}S^{\mu}+\partial_{\mu}A_{\nu}\frac{\partial S^{\mu}}{\partial A_{\nu}},
\end{eqnarray}
and designate
\begin{eqnarray}
\label{gauge-hibert-mom-def}
\pi^{\mu\nu}=\frac{\partial S^{\mu}}{\partial A_{\nu}},
\end{eqnarray}
then we derive the De Donder-Weyl equation for the Abelian gauge field
\begin{eqnarray}
\label{gauge-hibert-weyl-def}
\partial_{\mu}S^{\mu}+\mathscr{H}\left(A_{\mu},\pi^{\mu\nu}=\frac{\partial S^{\mu}}{\partial A_{\nu}}\right)=0.
\end{eqnarray}
For the covariant Hamiltonian (\ref{gauge-ham}), its corresponding De Donder-Weyl equation is
\begin{eqnarray}
\label{gauge-hibert-weyl}
\partial_{\mu}S^{\mu}+\frac{\Lambda}{2}\frac{\partial S^{\mu}}{\partial A_{\nu}}\frac{\partial S_{\mu}}{\partial A^{\nu}}
-\frac{\Lambda}{2}\theta\varepsilon^{\mu\nu\lambda}\frac{\partial S_{\mu}}{\partial A^{\nu}}A_{\lambda}+\frac{\Lambda}{4}\theta^2A_{\mu}A^{\mu}=0.
\end{eqnarray}

We employ the \lq\lq{embedding} method\rq\rq~to find solutions for the De Donder-Weyl equation (\ref{gauge-hibert-weyl}). At first, a solution of the gauge fixed Euler-Lagrange equation (\ref{gauge-ham-cano-euler}) can be given by
\begin{eqnarray}
\label{gauge-ham-cano-euler-sol}
B_{\nu}(x)&=&\left(-\frac{2}{5}\omega^2 C r+1\right)B_{\nu}(z)+{\omega}C\varepsilon_{\nu\alpha\beta}B^{\alpha}(z)(x^{\beta}-z^{\beta})
-\frac{2}{5}\omega^2 C B_{\alpha}(z)(x^{\alpha}-z^{\alpha})(x_{\nu}-z_{\nu}),\\
r&=&(x^{\mu}-z^{\mu})(x_{\mu}-z_{\mu}),~~~\omega=\Lambda\theta,\nonumber
\end{eqnarray}
where $z^{\mu}$ and $C$ are constant. We have used the initial condition $B_{\mu}(x)\vert_{z}=B_{\mu}(z)$ to normalize the solution.
Noticing that the De Donder-Weyl equation (\ref{gauge-hibert-weyl}) is quadratic, we can suppose a closed quadratic polynomial for $S^{\mu}$
\begin{eqnarray}
\label{gauge-hibert-weyl-supp}
S^{\mu}(x)=S^{\ast\mu}(x)+P^{\mu\nu}(x)\left[A_{\nu}(x)-B_{\nu}(x)\right]
+Y^{\mu\alpha\beta}(x)\left[A_{\alpha}(x)-B_{\alpha}(x)\right]\left[A_{\beta}(x)-B_{\beta}(x)\right].
\end{eqnarray}
Obviously, $Y^{\mu\alpha\beta}(x)$ should be symmetrical about its indices $\alpha$ and $\beta$. Substituting this expression
into eq.~(\ref{gauge-hibert-weyl}), we can get a quadratic polynomial of $\left[A_{\mu}(x)-B_{\mu}(x)\right]$. Then the
De Donder-Weyl equation (\ref{gauge-hibert-weyl}) can be satisfied by
\begin{eqnarray}
\label{gauge-hibert-weyl-supp-poly-0}
\partial_{\mu}S^{\ast\mu}(x)-\partial_{\mu}B_{\nu}(x)P^{\mu\nu}+\frac{\Lambda}{2}P^{\mu\nu}P_{\mu\nu}
-\frac{\Lambda}{2}\theta\varepsilon^{\mu\nu\lambda}P_{\mu\nu}B_{\lambda}(x)+\frac{\Lambda}{4}\theta^2B_{\mu}(x)B^{\mu}(x)&=&0,\\
\label{gauge-hibert-weyl-supp-poly-1}
\left[A_{\alpha}(x)-B_{\alpha}(x)\right]N^{\alpha}&=&0,\\
\label{gauge-hibert-weyl-supp-poly-2}
\left[A_{\alpha}(x)-B_{\alpha}(x)\right]\left[A_{\beta}(x)-B_{\beta}(x)\right]M^{\alpha\beta}&=&0,
\end{eqnarray}
where $N^{\alpha}$ and $M^{\alpha\beta}$ are defined by
\begin{eqnarray}
\label{gauge-hibert-weyl-supp-poly-1-def-n}
N^{\alpha}&=&\left(\partial_{\mu}P^{\mu\alpha}-\frac{\Lambda}{2}\theta\varepsilon^{\mu\nu\alpha}\pi_{\mu\nu}
+\frac{\Lambda}{2}\theta^2 B^{\alpha}(x)\right)
-2\left(\partial_{\mu}B_{\beta}(x)-\Lambda P_{\mu\beta}+\frac{\Lambda}{2}\theta\varepsilon_{\mu\beta\lambda}B^{\lambda}(x)\right)Y^{\mu\alpha\beta},\\
\label{gauge-hibert-weyl-supp-poly-2-def-m}
M^{\alpha\beta}&=&\partial_{\mu}Y^{\mu\alpha\beta}+2\Lambda \eta^{\beta\lambda}Y^{\mu\alpha\nu}Y_{\mu\lambda\nu}
-\Lambda\theta \eta^{\beta\lambda}\varepsilon^{\mu\nu\alpha}Y_{\mu\nu\lambda}+\frac{\Lambda}{4}\theta^2\eta^{\alpha\beta}.
\end{eqnarray}
As that in the \lq\lq{embedding} method\rq\rq, supposing the relations
\begin{eqnarray}
\label{gauge-ham-cano-x-supp}
\partial_{\mu}B_{\beta}(x)&=&\Lambda P_{\mu\beta}-\frac{\Lambda}{2}\theta\varepsilon_{\mu\beta\lambda}B^{\lambda}(x),\\
\label{gauge-ham-cano-p-supp}
\partial_{\mu}P^{\mu\alpha}&=&\frac{\Lambda}{2}\theta\varepsilon^{\mu\nu\alpha}\pi_{\mu\nu}
-\frac{\Lambda}{2}\theta^2 B^{\alpha}(x),
\end{eqnarray}
eq.~(\ref{gauge-hibert-weyl-supp-poly-1}) gets satisfied. Eqs.~(\ref{gauge-ham-cano-x-supp}) and (\ref{gauge-ham-cano-p-supp}) are self-consistent because they
are just the canonical Hamiltonian equations (\ref{gauge-ham-cano-x}) and (\ref{gauge-ham-cano-p}) and $B_{\mu}(x)$ solves the Euler-Lagrange
equation (\ref{gauge-ham-cano-euler}). While eq.~(\ref{gauge-hibert-weyl-supp-poly-0}) is transformed to be
\begin{eqnarray}
\label{gauge-hibert-weyl-supp-poly-0-trans}
\partial_{\mu}S^{\ast\mu}(x)=\partial_{\mu}B_{\nu}(x)P^{\mu\nu}-\frac{\Lambda}{2}P^{\mu\nu}P_{\mu\nu}
+\frac{\Lambda}{2}\theta\varepsilon^{\mu\nu\lambda}P_{\mu\nu}B_{\lambda}(x)-\frac{\Lambda}{4}\theta^2B_{\mu}(x)B^{\mu}(x)\nonumber\
=\mathscr{L}^{\ast}\Bigl(B_{\mu}(x)\Bigr).
\end{eqnarray}
The solutions for $P^{\mu\nu}$ can be obtained from eq.~(\ref{gauge-ham-cano-x-supp})
%\begin{widetext}
\begin{eqnarray}
\label{gauge-hibert-weyl-sol-1}
P_{\mu\nu}&=&\left[\frac{1}{2}+C-\frac{C^2}{10}\omega^2r\right]\theta\varepsilon_{\mu\nu\lambda}B^{\lambda}
+\frac{\theta}{10}{\omega}C B_{\mu}(z)(x_{\nu}-z_{\nu})-\frac{9\theta}{10}{\omega}C B_{\nu}(z)(x-z)_{\mu}\nonumber\\
&-&\frac{\theta}{5}{\omega^2}C B^{\alpha}(z)(x-z)_{\alpha}\varepsilon_{\mu\nu\lambda}(x-z)^{\lambda}
-\frac{2\theta}{5}{\omega}C B^{\alpha}(z)(x-z)_{\alpha}\eta_{\mu\nu}.
\end{eqnarray}
%\end{widetext}
The solution for $S^{\ast\mu}$ can be obtained from eq.~(\ref{gauge-hibert-weyl-supp-poly-0-trans})
%\begin{widetext}
\begin{eqnarray}
\label{gauge-hibert-weyl-sol-0}
S^{\ast\mu}(x)&=&\left[\frac{1}{3}\Bigl(C^2+C\Bigr)\omega\theta-\frac{1}{125}\theta\omega^3C^2r+C_1{r^{-\frac{3}{2}}}\right]B_{\alpha}(z)B^{\alpha}(z)(x-z)^{\mu}\\
&+&r^{-\frac{3}{2}}\exp^{-r}
\left[C_2+\frac{4}{25}\theta\omega^3C^2\int_0^{r}s^{\frac{1}{2}}\exp^{s}ds\right]\Bigl[B_{\alpha}(z)(x-z)^{\alpha}\Bigr]^2(x-z)^{\mu}.\nonumber
\end{eqnarray}
%\end{widetext}
While a solution of eq.~(\ref{gauge-hibert-weyl-supp-poly-2}) for $Y^{\mu\alpha\beta}(x)$ can be given by
\begin{eqnarray}
\label{gauge-hibert-weyl-sol-2}
Y^{\mu\alpha\beta}(x)=\frac{1}{2\Lambda}\frac{\omega}{\sqrt{2\sigma}}
\cot\left(\frac{\omega}{\sqrt{2\sigma}}k_{\lambda}(x-z)^{\lambda}\right)\eta^{\alpha\beta}k^{\mu},\hspace{3mm}\sigma=k_{\lambda}k^{\lambda},
\end{eqnarray}
where $k^{\mu}$ is a constant vector.

\subsection{Derivation of the Discretized Lagrangian}\label{sec:4.2}

Following the procedures in section \ref{sec:3.1.3}, we can derive the lattice Lagrangian from the solution (\ref{gauge-hibert-weyl-supp}) for the gauge field.
Because the solutions (\ref{gauge-hibert-weyl-sol-1}), (\ref{gauge-hibert-weyl-sol-0}) and (\ref{gauge-hibert-weyl-sol-2}) seem
complicated, the calculations will be laborious. A useful skill to perform these calculations is to employ the additive property of limiting, that is,
we calculate the limits term by term. Supposing the symmetrical lattice spacing $y^i-x^i=\epsilon$ and $\epsilon$ is small enough, we can get
\begin{eqnarray}
\label{latt-gauge-limit}
\hat{\mathscr{L}}_{\mathrm{lattice}}&=&\lim_{z^i{\rightarrow} x^i}\lim_{B_{\mu}(z^i){\rightarrow}A_{\mu}(x^i)}\tilde{\mathscr{L}}_{\mathrm{lattice}}
=\lim_{z^i{\rightarrow} x^i}\lim_{B_{\mu}(z^i){\rightarrow}A_{\mu}(x^i)} \sum_{i=0}^{2} \frac{S^{i}(y^{i})-S^{i}(x^{i})}{y^{i}-x^{i}}\\
&=&\frac{1}{2\Lambda}\sum_{i=0}^{2}\frac{\mathrm{sgn}(i)}{\epsilon^{2}}
\Bigl[A_{\alpha}(y^{i})-A_{\alpha}(x^{i})\Bigr]\Bigl[A^{\alpha}(y^{i})-A^{\alpha}(x^{i})\Bigr]
+\frac{\theta}{2}\sum_{i=0}^{2}\frac{1}{\epsilon}{\mathrm{sgn}(i)}\varepsilon^{i\nu\lambda}
\Bigl[A_{\nu}(y^{i})-A_{\nu}(x^{i})\Bigr]A_{\lambda}(x^{i})\nonumber\\
&+&\left(\frac{4}{5}\omega^{2}C-\frac{3\theta}{10}{\omega}C \right)\sum_{i=0}^{2}{\mathrm{sgn}(i)}\Bigl[A_{i}(y^{i})-A_{i}(x^{i})\Bigr]A_{i}(x^{i})\nonumber\\
&+&\left(\frac{4}{5}\omega^{2}C-\frac{9\theta}{10}{\omega}C\right) \sum_{i=0}^{2}\Bigl[A_{\alpha}(y^{i})-A_{\alpha}(x^{i})\Bigr]A^{\alpha}(x^{i}).\nonumber
\end{eqnarray}
The constant coefficients $C_1$ and $C_2$ in eq.~(\ref{gauge-hibert-weyl-sol-0}) have been settled to be zeros in order to avoid singularities in the limits.
For infinitesimal $\epsilon$, $A^{\alpha}(y^{i})-A^{\alpha}(x^{i}){\rightarrow}0$, we can also make the replacement
\begin{eqnarray}
\label{latt-gauge-limit-replace}
\frac{1}{\epsilon}\Bigl[A^{\alpha}(y^{i})-A^{\alpha}(x^{i})\Bigr]\xrightarrow{\epsilon{\rightarrow}0}\partial_{i}A^{\alpha}(x).
\end{eqnarray}
Then the lattice Lagrangian in (\ref{gauge-hibert-weyl-sol-1}) will approximate to the continuous one
\begin{eqnarray}
\label{latt-gauge-limit-replace-continuous}
\hat{\mathscr{L}}=\frac{1}{2\Lambda}\partial_{\mu}A_{\nu}\partial^{\mu}A^{\nu}+\frac{\theta}{2}\varepsilon^{\mu\nu\lambda}\partial_{\mu}A_{\nu}A_{\lambda},
\end{eqnarray}
which is just the gauge fixed Lagrangian (\ref{gauge-lag-gauge-fix}) with the gauge choice $\xi=\Lambda$.

From the foregoing discussions, it seems that we did not derive anything new from the solutions of the De Donder-Weyl equation. However, we caution that the
lattice Lagrangian (\ref{latt-gauge-limit}) is not the unique Lagrangian we can derive from the solutions of the De Donder-Weyl equation.
As we did in section \ref{sec:2.1.3}, we can derive another lattice Lagrangian from another solution of the De Donder-Weyl equation.
This is partially because eq.~(\ref{gauge-hibert-weyl-sol-2}) is not the only solution of eq.~(\ref{gauge-hibert-weyl-supp-poly-2}).
A second solution of eq.~(\ref{gauge-hibert-weyl-supp-poly-2}) can be given by
\begin{eqnarray}
\label{gauge-hibert-weyl-sol-2-second}
Y^{\mu\alpha\beta}(x)=-\frac{1}{2\Lambda}\frac{\omega}{\sqrt{2\sigma}}
\tan\left(\frac{\omega}{\sqrt{2\sigma}}k_{\lambda}(x-z)^{\lambda}\right)\eta^{\alpha\beta}k^{\mu}.\nonumber\\
\end{eqnarray}
Together with the solutions in eqs.~(\ref{gauge-hibert-weyl-sol-1}) and (\ref{gauge-hibert-weyl-sol-0}), eq.~(\ref{gauge-hibert-weyl-sol-2-second}) completes a new
solution for the De Donder-Weyl equation (\ref{gauge-hibert-weyl}). From this new solution, repeating the procedures as we did in deriving the lattice
Lagrangian (\ref{latt-gauge-limit}), we can derive a new lattice Lagrangian
\begin{eqnarray}
\label{latt-gauge-limit-massive}
\mathscr{L}_{\mathrm{lattice}}
&=&-{\omega}C^2{\theta}A_{\alpha}(x)A^{\alpha}(x)
-\frac{3\theta}{10}{\omega}C\sum_{i=0}^{2}{\mathrm{sgn}(i)}\Bigl[A_{i}(y^{i})-A_{i}(x^{i})\Bigr]A_{i}(x^{i})\\
&+&\left(\frac{1}{2}+C\right)\theta\sum_{i=0}^{2}\frac{1}{\epsilon}{\mathrm{sgn}(i)}\varepsilon^{i\nu\lambda}
\Bigl[A_{\nu}(y^{i})-A_{\nu}(x^{i})\Bigr]A_{\lambda}(x^{i})
-\frac{9\theta}{10}{\omega}C\sum_{i=0}^{2}\Bigl[A_{\alpha}(y^{i})-A_{\alpha}(x^{i})\Bigr]A^{\alpha}(x^{i}).\nonumber
\end{eqnarray}
This Lagrangian is similar to the lattice Lagrangian (\ref{latt-gauge-limit}). Actually, the Lagrangian (\ref{latt-gauge-limit-massive}) is a
part of the lattice Lagrangian (\ref{latt-gauge-limit}). What causes these differences is that the solution (\ref{gauge-hibert-weyl-sol-2}) contributes to the lattice Lagrangian
in the approximation of small enough $\epsilon$ while the solution (\ref{gauge-hibert-weyl-sol-2-second}) does not. For infinitesimal $\epsilon$, making the
replacement (\ref{latt-gauge-limit-replace}), we then derive the corresponding continuous one
\begin{eqnarray}
\label{latt-gauge-limit-massive-replace-continuous-2}
\mathscr{L}=\left(\frac{1}{2}+C\right)\theta\varepsilon^{\mu\nu\lambda}
\partial_{\mu}A_{\nu}(x)A_{\lambda}(x)-{\omega}C^2{\theta}A_{\alpha}(x)A^{\alpha}(x),\nonumber\\
\end{eqnarray}
where the corresponding Euler-Lagrangian equation will be
\begin{eqnarray}
\label{latt-gauge-limit-massive-replace-continuous-2-euler}
\left(\varepsilon^{\mu\nu\alpha}\partial_{\nu}-\frac{2C^2}{2C+1}\omega\eta^{\mu\alpha}\right)A_{\alpha}(x)=0.
\end{eqnarray}
Obviously, eq.~(\ref{latt-gauge-limit-massive-replace-continuous-2-euler}) is extremely similar to eq.~(\ref{gauge-lag-euler-dual}),
which is the equation of motion of the dual field strength ${}^{\ast}F^{\mu}$. Actually, when $C=1+\sqrt{3}$, $\frac{2C^2}{2C+1}=1$,
eq.~(\ref{gauge-lag-euler-dual}) is recovered up to a minus sign, which has no effect on the equation of motion of second order.
Eq.~(\ref{latt-gauge-limit-massive-replace-continuous-2-euler}) demonstrates the massive property of the gauge field as that
eq.~(\ref{gauge-lag-euler-dual}) does.

As in section \ref{sec:2.1.3}, we can find a third solution of eq.~(\ref{gauge-hibert-weyl-supp-poly-2}) for $Y^{\mu\alpha\beta}$
\begin{eqnarray}
\label{gauge-hibert-weyl-sol-2-third}
Y^{\mu\alpha\beta}(x)=-\frac{1}{2\Lambda}\frac{1}{r}
\left(1+\omega\sqrt{\frac{r}{2}}\tan\Bigl(\omega\sqrt{\frac{r}{2}}\Bigr)\right)\eta^{\alpha\beta}(x-z)^{\mu},\hspace{2mm}r=(x-z)^{\mu}(x-z)_{\mu}.
\end{eqnarray}
Associated with eqs.~(\ref{gauge-hibert-weyl-sol-1}) and (\ref{gauge-hibert-weyl-sol-0}), eq.~(\ref{gauge-hibert-weyl-sol-2-third}) constructs a third solution
for $S^{\mu}$. Following the similar limiting procedure above, we can derive the third Lagrangian
\begin{eqnarray}
\label{latt-gauge-limit-replace-continuous-3}
\hat{\mathscr{L}}=-\frac{1}{2\Lambda}\partial_{\mu}A_{\nu}\partial^{\mu}A^{\nu}+\left(\frac{1}{2}+2C\right)\theta\varepsilon^{\mu\nu\lambda}
\partial_{\mu}A_{\nu}(x)A_{\lambda}(x)-(1+2C)C{\omega}{\theta}A_{\alpha}(x)A^{\alpha}(x).
\end{eqnarray}
This Lagrangian has a mass term, but its kinetic has the minus sign, so this is also a ghost kinetic term. This is because the
solution (\ref{gauge-hibert-weyl-sol-2-third}) contributes to the Lagrangian opposite to that of the solution (\ref{gauge-hibert-weyl-sol-2}).

\section{Mass Generating Mechanism for a Fermion Field}\label{sec:5}

\subsection{De Donder-Weyl Theory for a Fermion Field}\label{sec:5.1}

We consider a massless fermion field in $D$ dimensional Minkowski space-time with the Lagrangian
\begin{eqnarray}
\label{fermion-lag}
\mathscr{L}=\frac{i}{2}\left[\bar{\psi}\gamma^{\mu}\partial_{\mu}\psi-\partial_{\mu}\bar{\psi}\gamma^{\mu}\psi\right]
-\frac{i}{\Lambda}\partial_{\mu}\bar{\psi}\sigma^{\mu\nu}\partial_{\nu}\psi,
\end{eqnarray}
where the Lorentzian metric $\eta=\mathrm{diag}(1,-1,-1,\cdots)$ is mostly negative diagonally, and $\sigma^{\mu\nu}=\frac{i}{2}[\gamma^{\mu},\gamma^{\nu}]$ is the
spin matrix. The last term of (\ref{fermion-lag}) is introduced in \cite{vonRieth:1982ac} to make the Legendre transformation regular\footnote{Of course,
we can also employ the Dirac-Bergmann method \cite{Dirac:1951zz,Dirac:1958sq,Bergmann:1949} to handle the problem of non-regularity.}.
The non-regularity of the Legendre transformation can also be cured by adding the Wilson's term $\partial_{\mu}\bar{\psi}\eta^{\mu\nu}\partial_{\nu}\psi$
\cite{Wilson:1974sk,Ginsparg:1981bj} or some more
general term as investigated in \cite{AngelesMartinez:2011nt}. However, an interesting feature
of the last term in (\ref{fermion-lag}) is that it is a total divergence
\begin{eqnarray}
\label{fermion-lag-term-surface}
\partial_{\mu}\bar{\psi}\sigma^{\mu\nu}\partial_{\nu}\psi=\partial_{\mu}\left[\bar{\psi}\sigma^{\mu\nu}\partial_{\nu}\psi\right],
\end{eqnarray}
so it has no effect on the equation of motion. The momentums are defined by the Legendre transformations
\begin{eqnarray}
\label{fermion-lag-lengendre-psi}
\bar{\pi}^{\mu}&=&
\frac{\partial\mathscr{L}}{\partial(\partial_{\mu}\psi)}=\frac{i}{2}\bar{\psi}\gamma^{\mu}-\frac{i}{\Lambda}\partial_{\alpha}\bar{\psi}\sigma^{\alpha\mu},\\
\label{fermion-lag-lengendre-psi-bar}
{\pi}^{\mu}&=&
\frac{\partial\mathscr{L}}{\partial(\partial_{\mu}\bar{\psi})}=-\frac{i}{2}\gamma^{\mu}\psi-\frac{i}{\Lambda}\sigma^{\mu\alpha}\partial_{\alpha}\psi,
\end{eqnarray}
which means the derivative of fields can be expressed with momentums
\begin{eqnarray}
\label{fermion-lag-lengendre-psi-inverse}
\partial_{\mu}\bar{\psi}&=&
\frac{i}{\Lambda}\left(\bar{\pi}^{\alpha}\tau_{\alpha\mu}-\frac{1}{2}\frac{1}{D-1}\bar{\psi}\gamma_{\mu}\right),\\
\label{fermion-lag-lengendre-psi-bar-inverse}
\partial_{\mu}\psi&=&
\frac{i}{\Lambda}\left(\tau_{\mu\alpha}{\pi}^{\alpha}+\frac{1}{2}\frac{1}{D-1}\gamma_{\mu}\psi\right),
\end{eqnarray}
where $D$ is the dimension of space-time, and $\tau_{\mu\nu}$ is defined by
\begin{eqnarray}
\label{fermion-lag-lengendre-tau-def}
\tau_{\mu\nu}=i\frac{D-2}{D-1}\eta_{\mu\nu}-\frac{1}{D-1}\sigma_{\mu\nu}.
\end{eqnarray}
The covariant Hamiltonian can be given by
\begin{eqnarray}
\label{fermion-lag-ham}
\mathscr{H}&=&{\bar{\pi}}^{\mu}\partial_{\mu}\psi+\partial_{\mu}{\bar{\psi}}{\pi}^{\mu}-\mathscr{L}\\
&=&\Lambda\left[i{\bar{\pi}}^{\mu}\tau_{\mu\nu}{\pi}^{\nu}+\frac{1}{4}\frac{D}{D-1}\bar{\psi}\psi\right]
+\Lambda\left[\frac{i}{2}\frac{1}{D-1}{\bar{\pi}}^{\mu}\gamma_{\mu}\psi-\frac{i}{2}\frac{1}{D-1}\bar{\psi}\gamma_{\mu}{\pi}^{\mu}\right].\nonumber
\end{eqnarray}
From this Hamiltonian, We can obtain the canonical equations of motion for $\psi$ and ${\pi}^{\mu}$
\begin{eqnarray}
\label{fermion-lag-ham-eq-psai}
\partial_{\mu}\psi&=&\frac{\partial\mathscr{H}}{\partial{\bar{\pi}}^{\mu}}=i\Lambda\tau_{\mu\alpha}{\pi}^{\alpha}+\frac{i}{2}\frac{1}{D-1}\gamma_{\mu}\psi,\\
\label{fermion-lag-ham-eq-pi}
-\partial_{\mu}{\pi}^{\mu}&=&\frac{\partial\mathscr{H}}{\partial\bar{\psi}}
=\Lambda\left(-\frac{i}{2}\frac{1}{D-1}\gamma_{\mu}{\pi}^{\mu}+\frac{1}{4}\frac{D}{D-1}\psi\right),
\end{eqnarray}
which together means that the Euler-Lagrangian equation is
\begin{eqnarray}
\label{fermion-lag-euler-psai}
i\gamma^{\mu}\partial_{\mu}\psi=0.
\end{eqnarray}
We notice that although a mass term appears in the Hamiltonian (\ref{fermion-lag-ham}), the Euler-Lagrangian equation is massless.
The contribution of the mass term is canceled secretly. For $\bar{\psi}$ and ${\bar{\pi}}^{\mu}$, the canonical equations of motion are
\begin{eqnarray}
\label{fermion-lag-ham-eq-psai-bar}
\partial_{\mu}\bar{\psi}&=&\frac{\partial\mathscr{H}}{\partial{\pi}^{\mu}}=
i\Lambda\bar{\pi}^{\alpha}\tau_{\alpha\mu}-\frac{i}{2}\frac{1}{D-1}\bar{\psi}\gamma_{\mu},\\
\label{fermion-lag-ham-eq-pi-bar}
-\partial_{\mu}{\bar{\pi}}^{\mu}&=&\frac{\partial\mathscr{H}}{\partial{\psi}}
=\Lambda\left(\frac{i}{2}\frac{1}{D-1}\bar{\pi}^{\mu}\gamma_{\mu}+\frac{1}{4}\frac{D}{D-1}\bar{\psi}\right).
\end{eqnarray}

Supposing the Lagrangian is the independent integral of Hilbert, we have
\begin{eqnarray}
\label{fermion-hilbert}
\mathscr{L}&=&\frac{d{S}^{\mu}}{d{x}^{\mu}}=\partial_{\mu}{S}^{\mu}+\partial_{\mu}\bar{\psi}\frac{\partial{S}^{\mu}}{\partial\bar{\psi}}
+\frac{\partial{S}^{\mu}}{\partial\psi}\partial_{\mu}{\psi}.
\end{eqnarray}
Furthermore, by designating
\begin{eqnarray}
\label{fermion-lag-ham-weyl-pi}
{\pi}^{\mu}&=&\frac{\partial{S}^{\mu}}{\partial\bar{\psi}},\\
\label{fermion-lag-ham-weyl-pi-bar}
\bar{\pi}^{\mu}&=&\frac{\partial{S}^{\mu}}{\partial{\psi}},
\end{eqnarray}
we then obtain the De Donder-Weyl equation for a fermion field
\begin{eqnarray}
\label{fermion-hilbert-weyl}
\partial_{\mu}{S}^{\mu}
+\mathscr{H}\left(\bar{\psi},\psi;{\pi}^{\mu}=\frac{\partial{S}^{\mu}}{\partial\bar{\psi}},\bar{\pi}^{\mu}=\frac{\partial{S}^{\mu}}{\partial{\psi}}\right)=0.
\end{eqnarray}
For the Hamiltonian (\ref{fermion-lag-ham}), its De Donder-Weyl equation is
\begin{eqnarray}
\label{fermion-weyl-donder}
\partial_{\mu}{S}^{\mu}
+\Lambda\left[i\frac{\partial{S}^{\mu}}{\partial{\psi}}\tau_{\mu\nu}\frac{\partial{S}^{\mu}}{\partial\bar{\psi}}
+\frac{1}{4}\frac{D}{D-1}\bar{\psi}\psi\right]+\Lambda\frac{i}{2}\frac{1}{D-1}\left(\frac{\partial{S}^{\mu}}{\partial{\psi}}\gamma_{\mu}\psi
-\bar{\psi}\gamma_{\mu}\frac{\partial{S}^{\mu}}{\partial\bar{\psi}}\right)=0.
\end{eqnarray}

We employ the \lq\lq{embedding} method\rq\rq~to seek solutions of the De Donder-Weyl equation (\ref{fermion-weyl-donder}). At the first step, a solution of the Euler-Lagrangian equation (\ref{fermion-lag-euler-psai}) is given by
\begin{eqnarray}
\label{fermion-euler-sol}
\chi(x)=\left(-\frac{DC}{\lambda}r\gamma_{\beta}k^{\beta}+1\right)\chi(z)+C\gamma_{\alpha}(x-z)^{\alpha}\chi(z),\hspace{2mm}
\lambda=k^{\alpha}k_{\alpha},\hspace{2mm}r=k_{\alpha}(x-z)^{\alpha},
\end{eqnarray}
where $k^{\alpha}$ and $z^{\alpha}$ are constant vectors, and $C$ is a constant of mass dimension. Here the initial designation $\chi(x)\vert_{z}=\chi(z)$ is employed.
Because eq.~(\ref{fermion-weyl-donder}) is quadratical, we can suppose a closed quadratical polynomial for ${S}^{\mu}$
\begin{eqnarray}
\label{fermion-weyl-donder-supp}
{S}^{\mu}(x)={S}^{\ast\mu}+\bar{P}^{\mu}\left[\psi(x)-\chi(x)\right]+\left[\bar{\psi}(x)-\bar{\chi}(x)\right]{P}^{\mu}
+\left[\bar{\psi}(x)-\bar{\chi}(x)\right]{R}^{\mu}\left[\psi(x)-\chi(x)\right].
\end{eqnarray}
Substituting (\ref{fermion-weyl-donder-supp}) into (\ref{fermion-weyl-donder}), we obtain a quadratical polynomials about $\left[\psi(x)-\chi(x)\right]$. The
De Donder-Weyl equation (\ref{fermion-weyl-donder}) can then be satisfied by supposing
%\begin{widetext}
\begin{eqnarray}
\label{fermion-weyl-donder-supp-poly-0}
\partial_{\mu}{S}^{\ast\mu}-{\bar{P}}^{\mu}\partial_{\mu}\chi(x)-\partial_{\mu}\bar{\chi}(x){P}^{\mu}&+&\hspace{0mm}\nonumber\\
\Lambda\left[i{\bar{P}}^{\mu}\tau_{\mu\nu}{P}^{\nu}+\frac{i}{2}\frac{1}{D-1}\left({\bar{P}}^{\mu}\gamma_{\mu}\chi(x)
-\bar{\chi}(x)\gamma_{\mu}{P}^{\mu}\right)+\frac{1}{4}\frac{D}{D-1}\bar{\chi}(x)\chi(x)\right]&=&0,\nonumber\\\\
\label{fermion-weyl-donder-supp-poly-1}
\left[\psi(x)-\chi(x)\right]:\hspace{1.5mm}
\partial_{\mu}{\bar{P}}^{\mu}-\partial_{\mu}\bar{\chi}(x){R}^{\mu}
+\Lambda\left[i{\bar{P}}^{\mu}\tau_{\mu\nu}{R}^{\nu}+\frac{i}{2}\frac{1}{D-1}\left({\bar{P}}^{\mu}\gamma_{\mu}
-\bar{\chi}(x)\gamma_{\mu}{R}^{\mu}\right)+\frac{1}{4}\frac{D}{D-1}\bar{\chi}(x)\right]&=&0,\nonumber\\\\
\label{fermion-weyl-donder-supp-poly-1-bar}
\left[\bar{\psi}(x)-\bar{\chi}(x)\right]:\hspace{1mm}
\partial_{\mu}{P}^{\mu}-{\bar{R}}^{\mu}\partial_{\mu}{\chi}(x)
+\Lambda\left[i{\bar{R}}^{\mu}\tau_{\mu\nu}{P}^{\nu}+\frac{i}{2}\frac{1}{D-1}\left({\bar{R}}^{\mu}\gamma_{\mu}\chi(x)
-\gamma_{\mu}{P}^{\mu}\right)+\frac{1}{4}\frac{D}{D-1}\chi(x)\right]&=&0,\nonumber\\\\
\label{fermion-weyl-donder-supp-poly-2}
\left[\bar{\psi}(x)-\bar{\chi}(x)\right]\left[\psi(x)-\chi(x)\right]:\hspace{13.5mm}
\partial_{\mu}{R}^{\mu}+\Lambda\left[i{\bar{R}}^{\mu}\tau_{\mu\nu}{R}^{\nu}
+\frac{i}{2}\frac{1}{D-1}\left({\bar{R}}^{\mu}\gamma_{\mu}-\gamma_{\mu}{R}^{\mu}\right)
+\frac{1}{4}\frac{D}{D-1}\right]&=&0.\nonumber\\
\end{eqnarray}
%\end{widetext}
Employing the \lq\lq{embedding} method\rq\rq, we can suppose that
\begin{eqnarray}
\label{fermion-weyl-donder-supp-poly-1-psi-bar}
\partial_{\mu}{\chi}(x)&=&
\Lambda\left[\tau_{\mu\nu}{P}^{\nu}+\frac{i}{2}\frac{1}{D-1}\gamma_{\mu}\chi(x)\right]\\
\label{fermion-weyl-donder-supp-poly-1-pi-bar}
-\partial_{\mu}{P}^{\mu}&=&\Lambda\left[-\frac{i}{2}\frac{1}{D-1}\gamma_{\mu}{P}^{\mu}+\frac{1}{4}\frac{D}{D-1}\chi(x)\right],
\end{eqnarray}
then eq.~(\ref{fermion-weyl-donder-supp-poly-1-bar}) can be satisfied. Eqs.~(\ref{fermion-weyl-donder-supp-poly-1-psi}) and
eq.~(\ref{fermion-weyl-donder-supp-poly-1-psi}) are just the Hamiltonian canonical equations (\ref{fermion-lag-ham-eq-psai}) and eq.~(\ref{fermion-lag-ham-eq-pi}).
Similarly, we can suppose
\begin{eqnarray}
\label{fermion-weyl-donder-supp-poly-1-psi}
\partial_{\mu}\bar{\chi}(x)&=&
\Lambda\left[i{\bar{P}}^{\alpha}\tau_{\alpha\mu}-\frac{i}{2}\frac{1}{D-1}\bar{\chi}(x)\gamma_{\mu}\right],\\
\label{fermion-weyl-donder-supp-poly-1-pi}
-\partial_{\mu}{\bar{P}}^{\mu}&=&
\Lambda\left[\frac{i}{2}\frac{1}{D-1}{\bar{P}}^{\mu}\gamma_{\mu}+\frac{1}{4}\frac{D}{D-1}\bar{\chi}(x)\right],
\end{eqnarray}
then eq.~(\ref{fermion-weyl-donder-supp-poly-1}) gets satisfied. They are just the Hermitian conjugate versions of
eqs.~(\ref{fermion-weyl-donder-supp-poly-1-psi-bar}) and (\ref{fermion-weyl-donder-supp-poly-1-pi-bar}).
Based on these assumptions, eq.~(\ref{fermion-weyl-donder-supp-poly-0}) is now transformed to be
\begin{eqnarray}
\label{fermion-weyl-donder-supp-poly-0-trans}
\partial_{\mu}{S}^{\ast\mu}&=&
-\Lambda\left[i{\bar{P}}^{\mu}\tau_{\mu\nu}{P}^{\nu}+\frac{1}{4}\frac{D}{D-1}\bar{\chi}(x)\chi(x)
+\frac{i}{2}\frac{1}{D-1}\left({\bar{P}}^{\mu}\gamma_{\mu}\chi(x)-\bar{\chi}(x)\gamma_{\mu}{P}^{\mu}\right)\right]\nonumber\\
&+&{\bar{P}}^{\mu}\partial_{\mu}\chi(x)+\partial_{\mu}\bar{\chi}(x){P}^{\mu}=\mathscr{L}^{\ast}\Bigl(\bar{\chi}(x),{\chi}(x)\Bigr).
\end{eqnarray}
Substituting the solution (\ref{fermion-euler-sol}) into (\ref{fermion-weyl-donder-supp-poly-1-psi-bar}) and (\ref{fermion-weyl-donder-supp-poly-1-psi}),
we can obtain solutions for $P^{\mu}$ and $\bar{P}^{\mu}$
\begin{eqnarray}
\label{fermion-weyl-donder-supp-poly-1-pi-sol}
{P}^{\mu}&=&\left[-\frac{i}{2}\gamma^{\mu}
+\frac{C}{\Lambda}\Bigl(\frac{D}{\lambda}\gamma_{\alpha}k^{\alpha}k^{\mu}-\gamma^{\mu}\Bigr)\right]\chi(z)
+\frac{i}{2}C\gamma^{\mu}\left[\frac{D}{\lambda}r\gamma_{\alpha}k^{\alpha}-\gamma_{\alpha}(x-z)^{\alpha}\right]\chi(z),\\
\label{fermion-weyl-donder-supp-poly-1-pi-sol-bar}
\bar{P}^{\mu}&=&\bar{\chi}(z)\left[\frac{i}{2}\gamma^{\mu}
+\frac{C}{\Lambda}\Bigl(\frac{D}{\lambda}\gamma_{\alpha}k^{\alpha}k^{\mu}-\gamma^{\mu}\Bigr)\right]
-\frac{i}{2}C\bar{\chi}(z)\left[\frac{D}{\lambda}r\gamma_{\alpha}k^{\alpha}-\gamma_{\alpha}(x-z)^{\alpha}\right]\gamma^{\mu}.
\end{eqnarray}
While the solution for ${S}^{\ast\mu}$ is given by
\begin{eqnarray}
\label{fermion-weyl-donder-supp-poly-0-trans-sol}
{S}^{\ast\mu}=\frac{C^2}{\Lambda\lambda}D(D-1)rk^{\mu}\bar{\chi}(z)\chi(z).
\end{eqnarray}
A solution for $R^{\mu}$ of eq.~(\ref{fermion-weyl-donder-supp-poly-2}) can be given by
\begin{eqnarray}
\label{fermion-weyl-donder-supp-poly-2-sol}
R^{\mu}&=&-\frac{1}{\Lambda}\frac{D-1}{D-2}\omega\tanh(\omega{r})k^{\mu},\hspace{4mm}D~{\neq}~2,\\
R^{\mu}&=&-\frac{1}{4}\frac{D}{D-1}{\Lambda}rk^{\mu},\hspace{17.25mm}D=2,
\end{eqnarray}
where $\omega^2=\frac{D(D-2)}{4\lambda}\Lambda^2$.

\subsection{Derivation of the Lattice Lagrangian for a Fermion field} \label{sec:5.2}

Following the limiting procedures, we can derive the lattice Lagrangian for the Fermion fields. Furthermore, we can derive the continuous Lagrangian as we did.
For simplicity, here we only give the continuous one
\begin{eqnarray}
\label{latt-fermion-continue}
\mathscr{L}=\frac{i}{2}\left[\bar{\psi}\gamma^{\mu}\partial_{\mu}\psi-\partial_{\mu}\bar{\psi}\gamma^{\mu}\psi\right]
+\frac{1}{\Lambda}D(D-1)C^2\bar{\psi}(x)\psi(x)
+\frac{1}{\Lambda}C\partial_{\mu}\left[\frac{D}{\lambda}\bar{\psi}\gamma_{\alpha}k^{\alpha}k^{\mu}\psi-\bar{\psi}\gamma^{\mu}\psi\right].
\end{eqnarray}
So we obtain a fermion Lagrangian with a mass term and some modifications. For $C=0$, we can get a conventional Lagrangian for massless fermions, but
for $C~{\neq}~0$, a mass term and modifications appear. We might imagine that we can recover the original Lagrangian (\ref{fermion-lag}) if
we employ other solutions of the De Donder-Weyl equation as we did for scalar fields and gauge fields. However, it seems it is very difficult to achieve this goal.
The reason is that the term $\partial_{\mu}\bar{\psi}\sigma^{\mu\nu}\partial_{\nu}\psi$ is difficult to be derived by the limiting procedure as we suggest.
As compensations for the loss of this term, a mass term and some modifications appear in the Lagrangian (\ref{latt-fermion-continue}). This situation
is quite similar to that in the topologically massive gauge theories. In that case, the kinetic term disappears while a mass term appears.

\section{SU(2) Yang-Mills Gauge Theory} \label{sec:6}

\subsection{De Donder-Weyl Equation of the Yang-Mills Gauge Theory} \label{sec:6.1}

For the $\mathrm{SU}(2)$ Yang-Mills gauge field in four dimensional Minkowski space-time, we consider the Lagrangian with gauge-fixed term
\begin{eqnarray}
\label{su2-lag}
\mathscr{L}=\frac{1}{4\Lambda}F^a_{\mu\nu}F^{a\mu\nu}+\frac{1}{2\xi}\partial_{\mu}A_{a\nu}\partial^{\nu}A_a^{\mu},\hspace{4mm}
F^a_{\mu\nu}=\partial_{\mu}A_{a\nu}-\partial_{\nu}A_{a\mu}+g \varepsilon^{abc}A_{b\mu}A_{c\nu},
\end{eqnarray}
where $\varepsilon^{abc}$ are structure constants of $\mathrm{SU}(2)$ Lie algebra. We introduce a constant $\Lambda$, which should be $-1$ for the four dimensional Lorentzian
metric $\eta=\mathrm{diag}(1,-1,-1,-1)$. The gauge fixed term is introduced in order to make the Legendre transformation regular.
The covariant Legendre transformation is given by
\begin{eqnarray}
\label{su2-lag-legendre}
\pi^{a\mu\nu}=\frac{\partial \mathscr{L}}{\partial(\partial_{\mu}A_{a\nu})}=\frac{1}{\Lambda}F^a_{\mu\nu}+\frac{1}{2\xi}\partial^{\nu}A_a^{\mu}.
\end{eqnarray}
Making the gauge choices $\xi=\Lambda$, the derivatives of the field can be expressed by the momentum
\begin{eqnarray}
\label{su2-lag-legendre-derivative}
\partial^{\mu}A_a^{\nu}=\Lambda\pi^{a\mu\nu}-g \varepsilon^{abc}A_{b}^{\mu}A_{c}^{\nu}.
\end{eqnarray}
Then the covariant Hamiltonian is found to be
\begin{eqnarray}
\label{su2-ham}
\mathscr{H}=\partial_{\mu}A_{\nu}\pi^{a\mu\nu}-\mathscr{L}
=\frac{\Lambda}{2}\pi^{a\mu\nu}\pi^a_{\mu\nu}-g\varepsilon^{abc}A_{b\mu}A_{c\nu}\pi^{a\mu\nu}
+\frac{g^2}{4\Lambda}\varepsilon^{abc}A_{b\mu}A_{c\nu}\varepsilon^{afh}A_{f}^{\mu}A_{h}^{\nu},
\end{eqnarray}
which yields the canonical equations of motion
\begin{eqnarray}
\label{su2-ham-x}
\partial_{\mu}A_{a\nu}&=&\frac{\partial \mathscr{H}}{\partial\pi^{a\mu\nu}}=\Lambda\pi_{a\mu\nu}-g\varepsilon^{abc}A_{b\mu}A_{c\nu},\\
\label{su2-ham-p}
-\partial_{\mu}\pi^{a\mu\nu}&=&\frac{\partial\mathscr{H}}{{\partial}A_{a\nu}}=
\frac{1}{\Lambda}\varepsilon^{dbc}A_{b\mu}A_{c\nu}\varepsilon^{dfa}A_{f}^{\mu}
+\varepsilon^{dba}A_{b\mu}(\pi^{d\mu\nu}-\pi^{d\nu\mu}).
\end{eqnarray}
Together they mean that the Euler-Lagrangian equation
\begin{eqnarray}
\label{su2-euler}
\partial_{\mu}\partial^{\mu}A_{a\nu}+g\varepsilon^{abc}\left(\partial_{\mu}A_b^{\mu}A_{c\nu}+2A_b^{\mu}\partial_{\mu}A_{c\nu}
-A_b^{\mu}\partial_{\nu}A_{c\mu}\right)+g^2\varepsilon^{abc}\varepsilon^{cdf}A_{b\mu}A_d^{\mu}A_f^{\nu}=0.
\end{eqnarray}
Supposing the Lagrangian to be the independent integral of Hilbert, we get
\begin{eqnarray}
\label{su2-hilbert}
\mathscr{L}=\frac{d S^{\mu}}{dx^{\mu}}=\partial_{\mu}S^{\mu}+\partial_{\mu}A_{a\nu}\frac{{\partial}S^{\mu}}{{\partial}A_{a\nu}}.
\end{eqnarray}
By designating
\begin{eqnarray}
\label{su2-hilbert-desig}
\pi^{a\mu\nu}=\frac{{\partial}S^{\mu}}{{\partial}A_{a\nu}},
\end{eqnarray}
we then obtain the De Donder-Weyl equation for $\mathrm{SU}(2)$ gauge theory
\begin{eqnarray}
\label{su2-weyl-dedonder}
\partial_{\mu}S^{\mu}+\mathscr{H}\left(A_{a\nu},\pi^{a\mu\nu}=\frac{{\partial}S^{\mu}}{{\partial}A_a^{\nu}}\right)=0.
\end{eqnarray}
For the Hamiltonian (\ref{su2-ham}), the De Donder-Weyl equation is
\begin{eqnarray}
\label{su2-weyl}
\partial_{\mu}S^{\mu}+\frac{\Lambda}{2}\frac{{\partial}S^{\mu}}{{\partial}A_a^{\nu}}\frac{{\partial}S_{\mu}}{{\partial}A_{a\nu}}
-g\varepsilon^{abc}A_{b\mu}A_{c\nu}\frac{{\partial}S^{\mu}}{{\partial}A_{a\nu}}+
\frac{g^2}{4\Lambda}\varepsilon^{abc}A_{b\mu}A_{c\nu}\varepsilon^{afh}A_{f}^{\mu}A_{h}^{\nu}=0.
\end{eqnarray}

We employ the \lq\lq{embedding} method\rq\rq~to seek the solutions for the De Donder-Weyl equation (\ref{su2-weyl}). At the first step, we need a solution for
the Euler-Lagrange equation (\ref{su2-euler}), one of which solutions can be given by
\begin{eqnarray}
\label{su2-euler-sol}
B_a^{\mu}(x)=\frac{\omega\sqrt{\sigma}}{\sqrt{2}g}\mathrm{JacobiCN}\left(r,\frac{1}{\sqrt{2}}\right)\varphi_a^{\mu},\hspace{2mm}
\sigma=k^{\mu}k_{\mu},\hspace{2mm}r={\omega}k^{\mu}(x-z)_{\mu}
\end{eqnarray}
where $\varphi_a^{\mu}$ is a constant tensor, which is restricted by the relations
\begin{eqnarray}
\label{su2-euler-sol-restr}
\varphi_a^{\mu}\varphi_{b\mu}=-\delta_{ab},~~~\varphi_a^{\mu}k_{\mu}=0,
\end{eqnarray}
where $\delta_{ab}$ is the Kronecker tensor. A solution of these restrictions can be given by
\begin{eqnarray}
\label{su2-euler-sol-restr-sol}
\varphi_a^{\mu}&=&\delta_{a}^{1}b^{\mu}+\delta_{a}^{2}c^{\mu}+\delta_{a}^{3}d^{\mu},~b^{\mu}b_{\mu}=-1,~c^{\mu}c_{\mu}=-1,~d^{\mu}d_{\mu}=-1,\\
b^{\mu}c_{\mu}&=&0,~b^{\mu}d_{\mu}=0,~c^{\mu}d_{\mu}=0,~k^{\mu}b_{\mu}=0,~k^{\mu}c_{\mu}=0,~k^{\mu}d_{\mu}=0.\nonumber
\end{eqnarray}
This solution has been listed in \cite{Fushchych:1997ks}. For more solutions of the $\mathrm{SU}(2)$ Yang-Mills equation, see \cite{Fushchych:1997ks,Actor:1979in}. We suppose a series solution for $S^{\mu}$ of the following type
%\begin{widetext}
\begin{eqnarray}
\label{su2-weyl-ansatz}
S^{\mu}&=&S^{\ast\mu}+P^{a\mu\alpha}\left[A_{a\alpha}(x)-B_{a\alpha}(x)\right]
+R_{ab}^{\mu\alpha\beta}\left[A_{a\alpha}(x)-B_{a\alpha}(x)\right]\left[A_{b\beta}(x)-B_{b\beta}(x)\right]\\
&+&K_{abc}^{\mu\alpha\beta\gamma}\left[A_{a\alpha}(x)-B_{a\alpha}(x)\right]\left[A_{b\beta}(x)-B_{b\beta}(x)\right]\left
[A_{c\gamma}(x)-B_{c\gamma}(x)\right]\nonumber\\
&+&M_{abcd}^{\mu\alpha\beta\gamma\lambda}\left[A_{a\alpha}(x)-B_{a\alpha}(x)\right]
\left[A_{b\beta}(x)-B_{b\beta}(x)\right]\left[A_{c\gamma}(x)-B_{c\gamma}(x)\right]\left[A_{d\lambda}(x)-B_{d\lambda}(x)\right]\nonumber\\
&+&N_{abcdf}^{\mu\alpha\beta\gamma\lambda\tau}\left[A_{a\alpha}(x)-B_{a\alpha}(x)\right]\left[A_{b\beta}(x)-B_{b\beta}(x)\right]
\left[A_{c\gamma}(x)-B_{c\gamma}(x)\right]\left[A_{d\lambda}(x)-B_{d\lambda}(x)\right]\left[A_{f\tau}(x)-B_{f\tau}(x)\right]\nonumber\\
&+&\cdots.\nonumber
\end{eqnarray}
%\end{widetext}
Substituting this series into eq.~(\ref{su2-weyl}), we can obtain a series about $[A_{a\alpha}(x)-B_{a\alpha}(x)]$. Eq.~(\ref{su2-weyl}) can be satisfied if we suppose the coefficients of this series to be zeros. For the term independent of fields, we have
\begin{eqnarray}
\label{su2-weyl-ansatz-0}
\partial_{\mu}S^{\ast\mu}=P^{a\mu\alpha}\partial_{\mu}B_{a\alpha}+g\varepsilon^{abc}B_{b\mu}B_{c\nu}P^{a\mu\nu}
-\left[\frac{\Lambda}{2}P^{a\mu\nu}P_{a\mu\nu}+\frac{g^2}{4\Lambda}\varepsilon^{abc}B_{b\mu}B_{c\nu}\varepsilon^{afh}B_{f}^{\mu}B_{h}^{\nu}\right].
\end{eqnarray}
For the term of power 1, we have
\begin{eqnarray}
\label{su2-weyl-ansatz-1}
\partial_{\mu}P^{b\mu\beta}-g\varepsilon^{acb}B_{c\mu}(P^{a\mu\beta}-P^{a\beta\mu})&+&\nonumber\\
\frac{g^2}{\Lambda}\varepsilon^{hbc}\varepsilon^{hdf}B_{c\nu}B_f^{\nu}B_d^{\beta}
-(\partial_{\mu}B_{a\alpha}-{\Lambda}P_{a\mu\alpha}-g\varepsilon^{abc}B_{b\mu}B_{c\alpha})(R_{ab}^{\mu\alpha\beta}+R_{ab}^{\mu\beta\alpha})&=&0.
\end{eqnarray}
For the term of power 2, we have
%\begin{widetext}
\begin{eqnarray}
\label{su2-weyl-ansatz-2}
0&=&\left[A_{b\beta}(x)-B_{b\beta}(x)\right]\left[A_{c\gamma}(x)-B_{c\gamma}(x)\right]T_{bc}^{\beta\gamma},\\
\label{su2-weyl-ansatz-2-def-co}
T_{bc}^{\beta\gamma}&=&\partial_{\mu}R_{bc}^{\mu\beta\gamma}
+(-\partial_{\mu}B_{a\alpha}+{\Lambda}P_{a\mu\alpha}+g\varepsilon^{abc}B_{b\mu}B_{c\alpha})
(K_{abc}^{\mu\alpha\beta\gamma}+K_{bac}^{\mu\beta\alpha\gamma}+K_{bca}^{\mu\beta\gamma\alpha})\\
&+&\frac{\Lambda}{2}(R_{ab}^{\mu\alpha\beta}+R_{ba}^{\mu\beta\alpha})(R_{ac\mu\alpha}^{\hspace{6mm}\gamma}+R_{ca\mu\beta}^{\hspace{6mm}\gamma})
-g\varepsilon^{abc}P^{a\beta\gamma}
-g\varepsilon^{adc}B_{d\mu}(R_{ab}^{\mu\gamma\beta}+R_{ba}^{\mu\beta\gamma})\nonumber\\
&-&g\varepsilon^{acd}B_{d\mu}(R_{ab}^{\gamma\mu\beta}+R_{ba}^{\gamma\beta\mu})
+\frac{g^2}{4\Lambda}\left[2(\varepsilon^{hbc}\varepsilon^{hdf}+\varepsilon^{hbf}\varepsilon^{hdc})B_{d}^{\beta}B_{f}^{\gamma}
+2\varepsilon^{hbd}\varepsilon^{hcf}B_{d\mu}B_{f}^{\mu}\eta^{\beta\gamma}\right].\nonumber
\end{eqnarray}
%\end{widetext}
For the term of power 3, we have
%\begin{widetext}
\begin{eqnarray}
\label{su2-weyl-ansatz-3}
0&=&\left[A_{b\beta}(x)-B_{b\beta}(x)\right]\left[A_{c\gamma}(x)-B_{c\gamma}(x)\right]
\left[A_{d\lambda}(x)-B_{d\lambda}(x)\right]Q_{bcd}^{\beta\gamma\lambda},\\
\label{su2-weyl-ansatz-3-def-co}
Q_{bcd}^{\beta\gamma\lambda}&=&\partial_{\mu}K_{bcd}^{\mu\beta\gamma\lambda}
+(-\partial_{\mu}B_{a\alpha}+{\Lambda}P_{a\mu\alpha}+g\varepsilon^{abc}B_{b\mu}B_{c\alpha})
(M_{abcd}^{\mu\alpha\beta\gamma\lambda}+M_{bacd}^{\mu\beta\alpha\gamma\lambda}+M_{bcad}^{\mu\beta\gamma\alpha\lambda}
+M_{bcda}^{\mu\beta\gamma\lambda\alpha})\nonumber\\
&+&\Lambda(K_{abc}^{\mu\nu\beta\gamma}+K_{bac}^{\mu\beta\nu\gamma}+K_{bca}^{\mu\beta\gamma\nu})
(R_{ad\mu\nu}^{\hspace{6mm}\lambda}+R_{da\mu\hspace{1mm}\nu}^{\hspace{4mm}\lambda})
-g\varepsilon^{abc}(R_{ad}^{\beta\gamma\lambda}+R_{da}^{\beta\lambda\gamma})\\
&-&g\varepsilon^{afd}B_{f\mu}(K_{abc}^{\mu\lambda\beta\gamma}+K_{bac}^{\mu\beta\lambda\gamma}+K_{bca}^{\mu\beta\gamma\lambda})
-g\varepsilon^{adf}B_{f\nu}(K_{abc}^{\lambda\nu\beta\gamma}+K_{bac}^{\lambda\beta\nu\gamma}+K_{bca}^{\lambda\beta\gamma\nu})\nonumber\\
&+&\frac{g^2}{4\Lambda}\left[\varepsilon^{hbc}\varepsilon^{hdf}B_{f}^{\gamma}\eta^{\beta\lambda}
+\varepsilon^{hfc}\varepsilon^{hdb}B_{f}^{\lambda}\eta^{\beta\gamma}+\varepsilon^{hbf}\varepsilon^{hdc}B_{f}^{\gamma}\eta^{\beta\lambda}
+\varepsilon^{hbc}\varepsilon^{hfd}B_{f}^{\beta}\eta^{\gamma\lambda}\right].\nonumber
\end{eqnarray}
%\end{widetext}
For the term of power 4, we have
%\begin{widetext}
\begin{eqnarray}
\label{su2-weyl-ansatz-4}
0&=&\left[A_{b\beta}(x)-B_{b\beta}(x)\right]\left[A_{c\gamma}(x)-B_{c\gamma}(x)\right]
\left[A_{d\lambda}(x)-B_{d\lambda}(x)\right]\left[A_{f\tau}(x)-B_{f\tau}(x)\right]X_{bcdf}^{\beta\gamma\lambda\tau},\\
\label{su2-weyl-ansatz-4-def-co}
X_{bcdf}^{\beta\gamma\lambda\tau}&=&\partial_{\mu}M_{bcdf}^{\mu\beta\gamma\lambda\tau}
+\frac{g^2}{4\Lambda}\varepsilon^{hbc}\varepsilon^{hdf}\eta^{\beta\lambda}\eta^{\gamma\tau}
-g\varepsilon^{adf}(K_{abc}^{\lambda\tau\beta\gamma}+K_{bac}^{\lambda\beta\tau\gamma}+K_{bca}^{\lambda\beta\gamma\tau})\\
&+&(-\partial_{\mu}B_{a\alpha}+{\Lambda}P_{a\mu\alpha}+g\varepsilon^{abc}B_{b\mu}B_{c\alpha})
(N_{abcdf}^{\mu\alpha\beta\gamma\lambda\tau}+N_{abcdf}^{\mu\beta\alpha\gamma\lambda\tau}+N_{abcdf}^{\mu\beta\gamma\alpha\lambda\tau}
+N_{abcdf}^{\mu\beta\gamma\lambda\alpha\tau}+N_{abcdf}^{\mu\beta\gamma\lambda\tau\alpha})\nonumber\\
&+&\frac{\Lambda}{2}(K_{abc}^{\mu\nu\beta\gamma}+K_{bac}^{\mu\beta\nu\gamma}+K_{bca}^{\mu\beta\gamma\nu})
(K_{adf\mu\nu}^{\hspace{7mm}\lambda\tau}+K_{daf\mu\hspace{1mm}\nu}^{\hspace{6mm}\lambda\hspace{1mm}\tau}
+K_{dfa\mu\hspace{2mm}\nu}^{\hspace{5.5mm}\lambda\tau})\nonumber\\
&+&\Lambda(M_{abcd}^{\mu\alpha\beta\gamma\lambda}+M_{bacd}^{\mu\beta\alpha\gamma\lambda}+M_{bcad}^{\mu\beta\gamma\alpha\lambda}
+M_{bcda}^{\mu\beta\gamma\lambda\alpha})
(R_{af\mu\nu}^{\hspace{6mm}\tau}+R_{fa\mu\hspace{1mm}\nu}^{\hspace{5mm}\tau})\nonumber\\
&-&g\varepsilon^{ahf}B_{h\mu}(M_{abcd}^{\mu\tau\beta\gamma\lambda}+M_{bacd}^{\mu\beta\tau\gamma\lambda}+M_{bcad}^{\mu\beta\gamma\tau\lambda}
+M_{bcda}^{\mu\beta\gamma\lambda\tau})\nonumber\\
&-&g\varepsilon^{afh}B_{h\nu}(M_{abcd}^{\tau\nu\beta\gamma\lambda}
M_{bacd}^{\tau\beta\nu\gamma\lambda}+M_{bcad}^{\tau\beta\gamma\nu\lambda}
+M_{bcda}^{\tau\beta\gamma\lambda\nu}).\nonumber
\end{eqnarray}
%\end{widetext}
For simplicity, we do not display the terms of higher power here.
As that in the \lq\lq{embedding} method\rq\rq, we can make some assumptions to solve these equations. We suppose that
\begin{eqnarray}
\label{su2-weyl-ansatz-1-supp-x}
\partial_{\mu}B_{a\alpha}&=&{\Lambda}P_{a\mu\alpha}-g\varepsilon^{abc}B_{b\mu}B_{c\alpha},\\
\label{su2-weyl-ansatz-1-supp-p}
-\partial_{\mu}P^{b\mu\beta}&=&-g\varepsilon^{acb}B_{c\mu}(P^{a\mu\beta}-P^{a\beta\mu})
+\frac{g^2}{\Lambda}\varepsilon^{hbc}\varepsilon^{hdf}B_{c\nu}B_f^{\nu}B_d^{\beta},
\end{eqnarray}
then eq.~(\ref{su2-weyl-ansatz-1}) can be satisfied. Eqs.~(\ref{su2-weyl-ansatz-1-supp-x}) and (\ref{su2-weyl-ansatz-1-supp-p}) are just the canonical equations
(\ref{su2-ham-x}) and (\ref{su2-ham-p}). They are self-consistent because $B_{a\alpha}$ is a solution of Euler-Lagrangian equation (\ref{su2-euler}). On these
assumptions, eqs.~(\ref{su2-weyl-ansatz-2}), (\ref{su2-weyl-ansatz-3}) and (\ref{su2-weyl-ansatz-4}) are simplified largely. While
eq.~(\ref{su2-weyl-ansatz-0}) is transformed to be
\begin{eqnarray}
\label{su2-weyl-ansatz-0-trans}
\partial_{\mu}S^{\ast\mu}=P^{a\mu\alpha}\partial_{\mu}B_{a\alpha}-\frac{\Lambda}{2}P^{a\mu\nu}P_{a\mu\nu}
+g\varepsilon^{abc}B_{b\mu}B_{c\nu}P^{a\mu\nu}
-\frac{g^2}{4\Lambda}\varepsilon^{abc}B_{b\mu}B_{c\nu}\varepsilon^{afh}B_{f}^{\mu}B_{h}^{\nu}
=\mathscr{L}^{\ast}\left(B_{a\alpha}(x)\right).\nonumber\\
\end{eqnarray}
The solutions for $S^{\ast\mu}$ and $P^{a\mu\nu}$ therefore can be obtained straightforwardly. However, the solution for $R_{ab}^{\mu\beta\gamma}$ is not
easily found. We can suppose the solution for $R_{ab}^{\mu\alpha\beta}$ of the type
\begin{eqnarray}
\label{su2-weyl-ansatz-2-ansatz}
R_{ab}^{\mu\beta\gamma}=X(r)\delta_{ab}\eta^{\alpha\beta}k^{\mu}
+R(r)\varepsilon_{abf}\left(\varphi_{f}^{\alpha}\eta^{\beta\mu}-\varphi_{f}^{\beta}\eta^{\alpha\mu}\right)
+Y(r)\varphi_{a}^{\alpha}\varphi_{b}^{\beta}k^{\mu}+Z(r)\varphi_{a}^{\beta}\varphi_{b}^{\alpha}k^{\mu}
+K(r)\delta_{ab}\varphi_{f}^{\alpha}\varphi_{f}^{\beta}k^{\mu}.\nonumber\\
\end{eqnarray}
Substituting this expression into (\ref{su2-weyl-ansatz-2}), and employing the relations in eq.~(\ref{su2-euler-sol-restr}), we can then
know that (\ref{su2-weyl-ansatz-2}) is satisfied if the following ODEs hold
%\begin{widetext}
\begin{eqnarray}
\label{su2-weyl-ansatz-2-ansatz-x}
{\omega}\sigma\frac{dX}{dr}+2\Lambda\left(X^2-2R^2\right)+4g\phi(r)R-\frac{g^2}{\Lambda}\phi^2(r)&=&0,\\
\label{su2-weyl-ansatz-2-ansatz-r}
{\omega}\sigma\frac{dR}{dr}+2\Lambda\left(Z+2X-K\right)R-g\phi(r)(Z+X-K)+\frac{g}{2\Lambda}\omega\frac{d\phi(r)}{dr}&=&0,\\
\label{su2-weyl-ansatz-2-ansatz-y}
{\omega}\sigma\frac{dY}{dr}+2\Lambda\left(-2Z+2X-2K-3Y\right)Y-2g\phi(r)R&=&0,\\
\label{su2-weyl-ansatz-2-ansatz-z}
{\omega}\sigma\frac{dZ}{dr}-4\Lambda\left(R^2+KZ-XZ\right)+\frac{g^2}{2\Lambda}\phi^2(r)&=&0,\\
\label{su2-weyl-ansatz-2-ansatz-k}
{\omega}\sigma\frac{dK}{dr}-2\Lambda\left(K^2-2R^2-2KX+Z^2\right)+2g\phi(r)R-\frac{g^2}{2\Lambda}\phi^2(r)&=&0,
\end{eqnarray}
%\end{widetext}
where $\phi(r)=\frac{\omega\sqrt{\sigma}}{\sqrt{2}g}\mathrm{JacobiCN}\left(r,\frac{1}{\sqrt{2}}\right)$.
The exact solutions of these ODEs are difficult to obtain.
However, from the discussions in sections \ref{sec:3} and \ref{sec:4}, we know that only the behavior of the solutions at small enough $r$ is important. So we
turn to seek solutions at small enough $r$. In the foregoing ODEs, we designate
\begin{eqnarray}
\label{su2-weyl-ansatz-2-ansatz-phi}
\phi(r)=\frac{\omega\sqrt{\sigma}}{\sqrt{2}g}\mathrm{JacobiCN}\left(r,\frac{1}{\sqrt{2}}\right)\xrightarrow[]{r\rightarrow{0}}\frac{\omega\sqrt{\sigma}}{\sqrt{2}g},
\end{eqnarray}
that is, we replace $\phi(r)$ with $\frac{\omega\sqrt{\sigma}}{\sqrt{2}g}$. Then we can obtain the solutions at small enough $r$
\begin{eqnarray}
\label{su2-weyl-ansatz-2-ansatz-x-sol}
X&=&\frac{\omega\sqrt{\sigma}}{2\Lambda}\coth(\frac{r}{\sqrt{\sigma}}),~~Z=-\frac{\omega\sqrt{\sigma}}{4\Lambda}\coth(\frac{r}{\sqrt{\sigma}}),\\
Y&=&-\frac{\omega\sqrt{\sigma}}{3\Lambda}\frac{1}{\sinh(2\frac{r}{\sqrt{\sigma}})+C_1\sinh^2(\frac{r}{\sqrt{\sigma}})},~~R=0,~~
K=\frac{\omega\sqrt{\sigma}}{4\Lambda}\coth(\frac{r}{\sqrt{\sigma}}),\nonumber
\end{eqnarray}
or we can get another solution
\begin{eqnarray}
\label{su2-weyl-ansatz-2-ansatz-x-sol-second}
X&=&\frac{\omega\sqrt{\sigma}}{2\Lambda}\tanh(\frac{r}{\sqrt{\sigma}}),~~Z=-\frac{\omega\sqrt{\sigma}}{4\Lambda}\tanh(\frac{r}{\sqrt{\sigma}}),\\
Y&=&-\frac{\omega\sqrt{\sigma}}{3\Lambda}\frac{1}{\sinh(2\frac{r}{\sqrt{\sigma}})+C_2\cosh^2(\frac{r}{\sqrt{\sigma}})},~~R=0,~~
K=\frac{\omega\sqrt{\sigma}}{4\Lambda}\tanh(\frac{r}{\sqrt{\sigma}}).\nonumber
\end{eqnarray}
These solutions have expressions similar to those that we obtained in sections \ref{sec:2} and \ref{sec:3}. The solutions for $K_{abc}^{\mu\alpha\beta\gamma}$
and $M_{abcd}^{\mu\alpha\beta\gamma\lambda}$ are more difficult to obtain. We expect that they might not contribute to the lattice Lagrangian in the infinitesimal
lattice spacing limit as happened in sections \ref{sec:2} and \ref{sec:3} for nonlinear theories.

\subsection{Derivation of the Lattice Lagrangian} \label{sec:6.2}

In this section, we plan to derive the lattice Lagrangian for the $\mathrm{SU}(2)$ Yang-Mills gauge theory. However, we are immediately met with problems. The problem is
the components of the solution (\ref{su2-euler-sol}) are subjected to the constraint condition (\ref{su2-euler-sol-restr}). If we
take the limits $A_{a\mu}(x){\rightarrow}B_{a\mu}(z)$, then the components of $A_{a\mu}(x)$ must also be subjected to the constraint
condition (\ref{su2-euler-sol-restr}) in order to avoid singularities in the limits. Therefore, what we can derive by the foregoing limiting procedures
is a Lagrangian with its field components subjected to be the constraint condition (\ref{su2-euler-sol-restr}). An approach to overcome this problem is to
find more general solutions of the Euler-Lagrangian equation (\ref{su2-euler}). This means we need find solutions with more general initial conditions.
The initial value problem of the Yang-Mills equation has been discussed in \cite{Gu:1975df} based on theories of partial differential equations, and also can be addressed in the De Donder-Weyl theories as it has been did in \cite{Bruno:2010nv}. So it seems we have no problems in finding solutions of more general initial conditions. However, even if we can obtain a general solution for the Euler-Lagrangian equation, it is still a challenge to find a manifestly Lorentz covariant solution of the De Donde-Weyl equation, due to the difficulty in solving tensor equations. From our discussions on the restricted case, it seems that it is possible that we can obtain several kinds of solutions of the De Donde-Weyl equation.

In section \ref{sec:4}, we showed that we can derive a massive Lagrangian of first order for the Abelian topologically massive gauge theory. Furthermore, for the non-Abelian Yang-Mills theory, the massive infrared behavior of the gluon propagator has been observed in large volume lattice simulations \cite{Bogolubsky:2007ud,Cucchieri:2007md,Sternbeck:2007ug}. So an interesting question is whether we can also derive a massive Lagrangian for $\mathrm{SU}(2)$ Yang-Mills theory in our present framework. The answer is inconclusive. The derivation of lattice Lagrangians depends on the solutions of the De Donder-Weyl equation. We can speculate that there might exist a solution which would lead to a massive Lagrangian for $\mathrm{SU}(2)$ Yang-Mills, but we did not find such a solution so far because of the complexities of tensor equations.  The solution we can get is a restricted case as we showed in the last section. To derive a massive Lagrangian or  to develop some qualitative criteria which judge whether we can derive a massive Lagrangian is an interesting question which deserves our future considerations. However, our present framework can still help us to understand nonperturbative behaviors of Yang-Mills theory from another perspective. In \cite{Faddeev:1996zj,Faddeev:1998eq,Faddeev:2006sw}, the authors proposed that the $\mathrm{O}(3)$ nonlinear $\sigma$ model
\begin{eqnarray}
\label{su2-faddeev}
\mathcal{L}_{\mathrm{eff}}=\frac{m^2}{2}(\partial_{\mu}\bf{n})^2+\frac{1}{4}(\bf{n}\cdot\partial_{\mu}\bf{n}\times\partial_{\nu}\bf{n})^2+V(\bf{n}),\hspace{2mm}
\end{eqnarray}
where $\bf{n}$ is a three dimensional unit vector, could be related to the effective low energy Lagrangian of a $SU(2)$ Yang-Mills theory, because the Lagrangian (\ref{su2-faddeev}) supports solutions of closed knotted solitons \cite{Battye:1998pe,Hietarinta:1998kt,Lin04}. In \cite{Cho:2001fr}, supposing a special gauge field decomposition
\begin{eqnarray}
\label{su2-faddeev-decomp}
\bf{A}_{\mu}=C_{\mu}\bf{n}+\partial_{\mu}\bf{n}\times\bf{n}+\rho\partial_{\mu}\bf{n}+\sigma\partial_{\mu}\bf{n}\times\bf{n},
\end{eqnarray}
the authors show that an effective Lagrangian similar to (\ref{su2-faddeev}) can be derived from the Yang-Mills action after one loop functional integration. In our present framework, we can try to find a solution by supposing the gauge field expansion (\ref{su2-faddeev-decomp}), then by employing the~\lq\lq{embedding} method\rq\rq~and the limiting procedures, we can speculate that we can also derive a Lagrangian similar to (\ref{su2-faddeev}). Therefore, our present framework may provide another approach to understand the infrared behavior of Yang-Mills theories, although it is difficult to implement.

\section{Further Discussions} \label{sec:7}

\subsection{Discussions on Solution Dependence} \label{sec:7.1}

The main points of our paper can be summarized as follows. Starting with solutions of the Hamilton-Jacobi equation or the De Donder-Weyl equation, we can derive a
lattice Lagrangian by the limiting procedures introduced in sections \ref{sec:2} and \ref{sec:3}. The derived lattice Lagrangian can be employed to
formulate a lattice definition of path integrals. These steps make it possible to derive quantum theories from solutions of the Hamilton-Jacobi equation\footnote{For conventional lattice gauge theories, the lattice Lagrangian employs the gauge link variables so the problem of gauge invariance is resolved. While our derived
lattice Lagrangian depends on the field variables directly, so the gauge invariance is broken and we need introduce ghost fields to cure the problem
of renormalization.}. Here problems arise because we can obtain several kinds of solutions of the Hamilton-Jacobi equation~(or De Donder-Weyl equation for field
theories) generally. We can derive several different lattice Lagrangians from these different kinds of solutions. Some Lagrangians have the problem of stability, as we obtained ghost Lagrangians in sections \ref{sec:3} and \ref{sec:4}. There exist natural criteria for mechanics to select out the physical Lagrangian but they do not apply to field theories.

For the harmonic oscillator, we obtain 3 solutions (\ref{osci-hj-solution-poly}), (\ref{osci-hj-solution-poly-1}) and (\ref{osci-hj-solution-poly-non}) for its
Hamilton-Jacobi equation. The solution (\ref{osci-hj-solution-poly}) leads to the conventional Lagrangian by our limiting procedures, while solutions
(\ref{osci-hj-solution-poly-1}) and (\ref{osci-hj-solution-poly-non}) do not. However, we can use boundary conditions to fix the solutions. For a time evolution
from $t_a$ to $t_b$, we have a orbit from $(t_a,x_a)$ to $(t_b,x_b)$ in the configuration space. The classical action along this orbit can be evaluated, which is given by
\begin{eqnarray}
\label{osi-classic-action}
S(x_b,t_b;x_a,t_a)=\frac{m}{2}\frac{\omega}{\sin{\omega(t_b-t_a)}}[\cos{\omega(t_b-t_a)}(x^2_b+x^2_a)-2 x_b x_a].
\end{eqnarray}
By designating the constants of integral appropriately, solutions
(\ref{osci-hj-solution-poly-1}) and (\ref{osci-hj-solution-poly-non}) can both lead to classical actions equivalent to (\ref{osi-classic-action}).
So the uniqueness of the problem of mechanics can be resolved with the help of the classical action. However, this method cannot be applied to field theories, obviously.
The action of field depends on the volume of space-time. There is no unique definition for the action of field between two different space-time points. We might expect the integrability conditions introduced in section \ref{sec:2.1.1} to restrict the solutions of the De Donder-Weyl equation, but they cannot remove the nonphysical solutions. At least they do not work for the free scalar field theory. Another possible criterion is the requirement of employing a \lq\lq{complete} solution\rq\rq~of the De Donder-Weyl equation to derive the lattice Lagrangian. A \lq\lq{complete} solution\rq\rq~should depend on the $n\times{d}$ constants of the integral, where $n$ is the number of real components of field variables and $d$ is the number of space-time dimensions. But this criterion does not apply to the free scalar field theory neither, and there are also difficulties in finding \lq\lq{complete} solutions\rq\rq~for other field theories. The solutions for the gauge field and fermion field given in previous sections are not \lq\lq{complete} solutions\rq\rq. These problems imply our procedures are not complete and we need additional conditions to remove the nonphysical solutions. However, as we just discussed, it seems no appropriate criteria emerge naturally.

\subsection{comparisons with conventional lattice gauge theories} \label{sec:7.2}

Another problem we should mention is how to implement gauge symmetries in our derivation of lattice Lagrangians. In conventional lattice gauge theories, the gauge link
variables $U_{\mu}(x)$ are employed; therefore, the gauge invariance is manifest in appropriately chosen lattice Lagrangians. In our present framework, we meet the same problems that we usually meet in the framework of canonical quantization: The Lagrangian is not regular due to the gauge invariance. In this paper, we solve this problem by adding gauge fixed terms. The final Lagrangians we derived are gauge fixed Lagrangians but not gauge invariant ones. This is an unsatisfactory aspect of our framework compared to the conventional lattice gauge theories. It is a challenge to figure out whether we can implement gauge invariance by appropriate modifications in our present approach. Leaving the gauge symmetry problem aside, it seems that our approach has advantages on dealing with infrared behaviors of field theories. In section \ref{sec:4}, we derived a massive Lagrangian for a topologically massive gauge theories. In section \ref{sec:5}, we suggested a mass generating mechanism for fermions. In section \ref{sec:6}, we discussed the possibilities of deriving an effective Lagrangian for a Yang-Mills theory. These advantages are achieved by employing special solutions of De Donder-Weyl equations cleverly. However, we should caution that these advantages are not ensured, because we can always derive physical Lagrangians  associated with nonphysical ones while we do not have any natural criteria to select out physical ones. These aforementioned potential advantages could be verified until we can achieve a better understanding about the problem of solution dependence as we discussed in the last subsection.

\section{Conclusions} \label{sec:8}

The construction of covariant Hamilton-Jacobi theories or De Donder-Weyl theories can be traced back to the efforts to formulate
the Lorentz covariant canonical quantization from the Dirac canonical bracket\footnote{For several suggestions recently, we refer to
\cite{Kanatchikov:2002yh,Kanatchikov:2008he,Nikolic:2006fm,Ozaki:2005xn}. The De Donder-Weyl equation for gravity of Einstein-Hilbert action has been derived
in \cite{Horava:1990ba}. The application of De Donder-Weyl theories to canonical quantum gravity has been suggested in \cite{Kanatchikov:2000jz,Rovelli:2002hb}.}. These efforts met problems partially because the covariant Hamiltonian does not correspond to the total energy of the system. Rather than attempting to construct a covariant canonical formulation, we suggest deriving lattice Lagrangians from solutions of De Donder-Weyl equations by appropriate limiting procedures. It turns out we can obtain several different sectors from different kinds of solutions generally. We can obtain a massive Lagrangian of first order for a topologically massive gauge theory. We also find a surface term that can produce masses for fermions, which is a mass generating mechanism similar to that of topologically massive gauge theories. The problems are that we also obtain ghost Lagrangians, which means we might need an mechanism to get rid of these nonphysical results. For nonlinear theories like Yang-Mills theories, our analysis on restricted solutions suggest we can derive nontrivial lattice Lagrangians as we did for topologically massive gauge theories, but these remain speculative until we can handle complicated tensor equations.

\acknowledgments

We thank S. Kovalenko for useful discussions. This work is supported by Project Basal under Contract No. FB0821~(Chile).

\bibliographystyle{utphys}
\bibliography{reference}

\end{document}